\documentclass[manuscript]{aastex}

\usepackage{amsmath}
\usepackage{color}
\def\simleq{\; \raise0.3ex\hbox{$<$\kern-0.75em \raise-1.1ex\hbox{$\sim$}}\; }

\shorttitle{Dynamics of Three-Planet Systems}
\shortauthors{Shikita et al.}

\begin{document}


\title{The Dynamics of Three-Planet Systems: an Approach from Dynamical System}


\author{Bungo Shikita\altaffilmark{1}, Hiroko Koyama\altaffilmark{2,3},\& Shoichi Yamada\altaffilmark{1,3}}
\altaffiltext{1}{Science and Engineering, Waseda University, 3-4-1 Okubo, Shinjuku, Tokyo 169-8555, Japan}
\altaffiltext{2}{Department of Physics, Nagoya University, Nagoya, 464-8602, Japan}
\altaffiltext{3}{Advanced Research Institute for Science and Engineering, Waseda University, 3-4-1 Okubo, Shinjuku, Tokyo 169-8555, Japan}

\email{shikita@heap.phys.waseda.ac.jp}


\begin{abstract}

We study in detail the motions of three planets interacting with each
other under the influence of a central star. 
It is known that the system with more than two planets becomes
unstable after remaining quasi-stable for long times, leading to highly
eccentric orbital motions or ejections of some of the planets. In this paper,
we are concerned with the underlying physics for this quasi-stability
as well as the subsequent instability and advocate the so-called
"stagnant motion" in the phase space, which has been explored in the
field of dynamical system. We employ the Lyapunov exponent, the power
spectra of orbital elements and the distribution of the durations of 
quasi-stable motions to analyze the phase space structure of the
three-planet system, the simplest and hopefully representative one
that shows the instability. We find from the Lyapunov exponent that
the system is almost non-chaotic in the initial quasi-stable state
whereas it becomes intermittently chaotic thereafter. 
The non-chaotic motions produce the horizontal dense band in 
the action-angle plot whereas the voids correspond to the chaotic
motions. We obtain power laws for the power spectra of orbital eccentricities. 
Power-law distributions are also found for the durations of
quasi-stable states. All these results combined together, we may
reach the following picture: the phase space consists of the so-called KAM
tori surrounded by satellite tori and imbedded in the chaotic sea. 
The satellite tori have a self-similar distribution and are responsible
for the scale-free power-law distributions of the duration times. 
The system is trapped around one of the KAM torus and the satellites 
for a long time (the stagnant motion) and moves to another KAM torus
with its own satellites from time to time, corresponding to the
intermittent chaotic behaviors. 

\end{abstract}


\section{Introduction}

More than $300$ exoplanets have been discovered so far 
\footnote{http://exoplanet.eu for the latest information.} and,
interestingly, some of them have a quite different appearance from
that of our solar system. The existence of so-called "eccentric
planets", that is, planets with high orbital eccentricities, for
example, attracts attentions of many researchers \citep{obs}.
Stimulated by these observations, the planetary formation theory
including the origin of the eccentric planets has made substantial
progress over the years \citep{kki06}. In the standard theory,
terrestrial planets are thought to be formed through giant impacts of
proto-planets (planets with sub-Earth masses) in crossing orbits \citep{cw98}.
 Since N-body simulations suggest that  
proto-planets are formed in nearly circular orbits separated by
several Hill radii \citep{ki95}, some destabilizing processes
are expected to operate and make proto-planets originally in the 
well-separated circular orbits collide with each other and 
grow up to terrestrial planets.
 
The stability of the system with two planets around a central star 
has been thoroughly investigated in celestial mechanics and it is
known that there exists a critical orbital separation between the 
planets, beyond which the planets never experience close encounters
and the system remains stable forever \citep[]{mb80,g93}.
 The situation changes drastically, however, if another planet
is added to the system. Using numerical simulations, \citet{cwb96}
 demonstrated that the systems with more than two planets become
unstable even for large orbital separations. Although the planetary motions
remain regular for some time at first, one of the planets eventually
comes close enough to another (that is, within the Hill radius of the
latter), leading to subsequent orbital crossings.

This is a good news for the terrestrial formation theory and may also
account for the formation of the eccentric planets. In fact, \citet{mw02}
 numerically integrated the motions of three
Jupiter-mass planets and found in most of their simulations that 
one of the planets is ejected from the system and the others are left 
in the system with high eccentricities. Several attempts \citep[]{jt07,fr07}
 have been made to reproduce
the observed eccentricity distribution by the orbital
instability. Their results seem to be consistent with the observations
although the latter itself may be somewhat biased \citep{st08}.

The planetary motions in these systems are interesting in their own
right. As mentioned above, we commonly observe a long period of
quasi-regular motions that look like independent Keplerian motions 
before the eventual orbital crossings. The switch is sudden and quick.  
The duration of the quasi-regular motions is sensitive to the initial
orbital separations and eccentricities \citep[]{cwb96,ykm99}.
 Using numerical simulations and simplified
analytical models, \citet{zls07} claimed that the gradual
deviation from the Keplerian motions can be regarded as a random
walk process. These efforts notwithstanding, the underlying physics
behind the phenomena such as the long period of quasi-regular motions
followed by the sudden transition to chaotic states is remaining to be
revealed and is the main concern of this paper.

We attempt to understand this phenomenon as the so-called 
``stagnant motion'' in the phase space, which will be described below.
We pay attention to a similar phenomenon known in the field of
dynamical system. In nearly integrable systems, a sudden transition
from a regular motion sustained for a long time to chaotic motions is 
often observed. In the nonlinear lattice problems, for example, 
\citet{hs68} found that an initially imposed normal mode 
experiences sudden energy exchanges among several other
modes after long regular oscillations. They called this 
"the induction phenomenon" and referred to the duration of the regular 
motion as "the induction period" \citep{soah70}. 

\begin{figure}
\begin{center}
\resizebox{140mm}{!}{\plotone{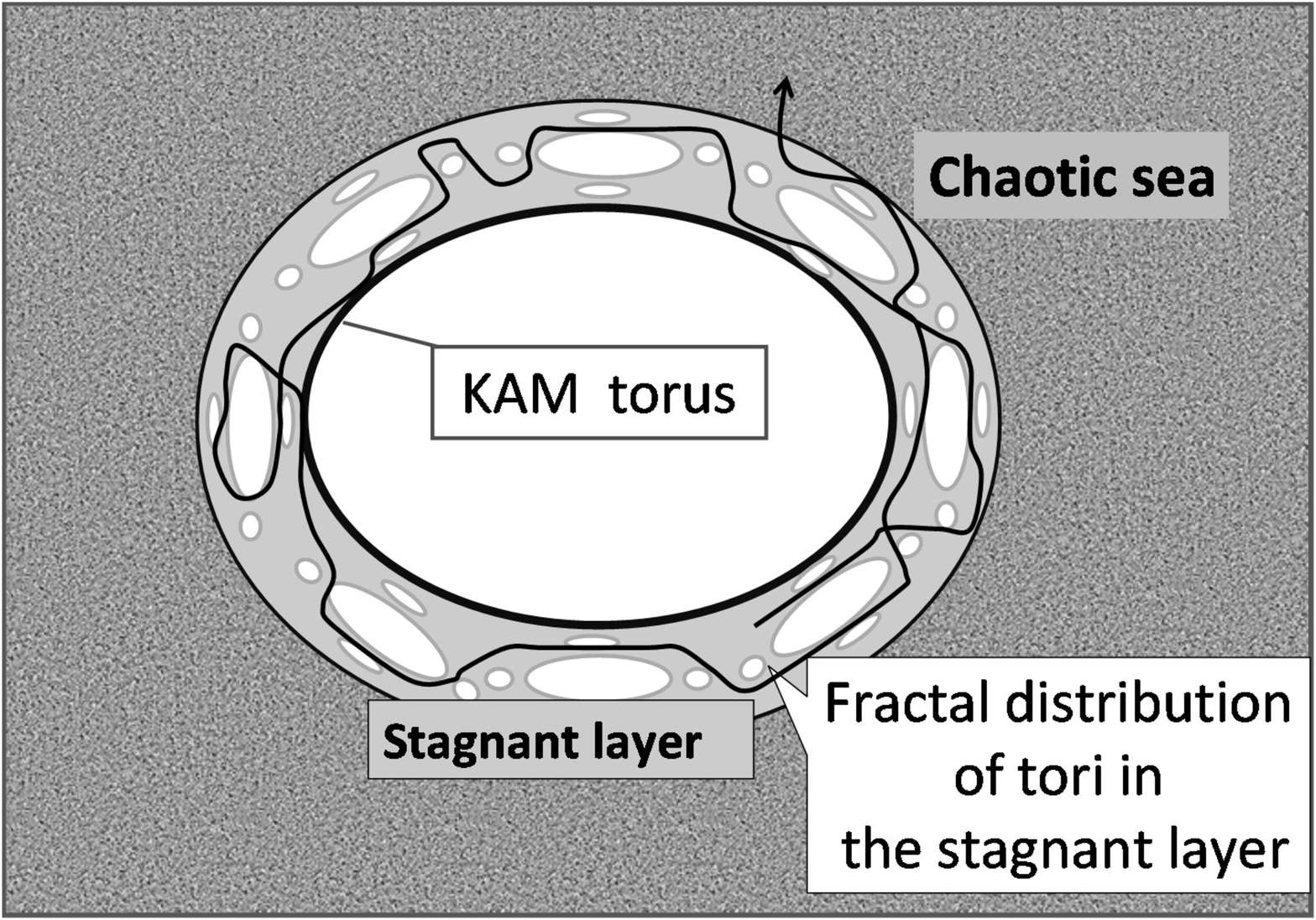}} \\
\caption{A schematic illustration of the structure in the phase space
  proposed by the stagnant motion model. The KAM torus is surrounded
  by many smaller tori distributed in a self-similar manner in the
  stagnant layer. An exemplary evolution in the phase space is 
  shown by a curve with an arrow. The system is trapped by the tori in 
  the stagnant layer and stay there for a long time, showing
  quasi-regular motions, which are turned into chaotic motions when
  the phase-space orbit escapes out of the stagnant layer into the
  chaotic sea.}
    \label{schematic1}	
\end{center}	
\end{figure}

\citet{aizawa} constructed a so-called "stagnant motion" model 
for this induction phenomenon. According to the KAM theorem 
\citep[]{kolmogorov,moser,arnold}, the phase space of a 
nearly integrable system retains tori, which exist in the integrable
system, if perturbations to the integrable system are
sufficiently weak. It is generally expected that the so-called KAM
tori will survive even for not so small perturbations. In the stagnant 
motion model, it is assumed that the KAM torus exists
in "the chaotic sea", the region corresponding to chaotic motions 
of the system, being surrounded by a thin layer called 
"the stagnant layer", in which
smaller tori are distributed in a self-similar manner 
(see Fig.~\ref{schematic1}).  
The system shows nearly regular behavior when the phase space orbit 
is trapped in the stagnant layer whereas the sudden transition to
chaotic motions occur when the orbit escapes out of the layer. This
model is successful in reproducing the scale-free power spectra and the
distribution of the induction period. In this paper we show some 
evidence to support the interpretation of the motions of three-planet
system as one of the induction phenomena and attempt to understand them 
in the frame work of the stagnant motion model.

The organization of the paper is as follows. 
We summarize the numerical models in section 2. 
In section 3, we describe the methods of analysis. 
The results are presented in section 4 and the summary and discussions 
are given in section 5. 

%%%%%%%%%%%%%%%%%%%%%%%%%%%%%%%%%%%%%%%%%%%%%%%%%%%%%%%%%%%%%%%%%%%%%%%%%%%%%%%%%%%%%%%
\section{Models}
%%%%%%%%%%%%%%%%%%%%%%%%%%%%%%%%%%%%%%%%%%%%%%%%%%%%%%%%%%%%%%%%%%%%%%%%%%%%%%%%%%%%%%%
In this paper, we restrict the investigation to \textit{the simplest} 
multi-planet system, which shows the behavior mentioned above: the
system consists of a central star with $M_* = 1 M_\odot$ and three
planets with an identical mass in coplanar orbits. We consider two
cases for the planetary mass, $m_{pl}$: (1) $m_{pl}=10^{-7}M_\odot$ 
(the proto-planet system) and (2) $m_{pl}=10^{-3}M_\odot$ (the Jupiter 
system). The initial semi-major axis of the innermost planet is set
to be $a_1=1$AU for the proto-planet system and $a_1=5$AU for the 
Jupiter system. Following \citet{cwb96}, we give the initial
radial locations of the outer planets as
\begin{equation}
a_{i+1}=a_{i}+\Delta r_{h(i,i+1)}, 
\label{eq:initial}
\end{equation}
where $a_i$ is the semi-major axis of the $i$-th planet counted from
the innermost one, and $r_{h(i,i+1)}$ is the Hill radius for the pair
of $i$-th and $i+1$-th planets defined as
\begin{equation}
r_{h(i,j)}= \biggl( \frac{m_i+m_j}{3M_*} \biggr)  ^{1/3}\frac{a_i+a_j}{2}.
\label{eq:hill}
\end{equation}
In addition, we also consider the case with $m_{pl}=10^{-3}M_\odot$, 
$a_1=5$AU, $a_2=7.25$AU, $a_3=9.5$AU, that is, the same parameter set
as that used in \citet{mw02} except for no
inclination in our model. 

The Hamiltonian in the barycentric coordinates consists of three parts
and is given as
\begin{eqnarray}
H & = & H_{Kep} + H_{*} + H_{int}, \label{eq:hamiltonian}\\
H_{Kep} & = & \sum _{i = 1}^n \biggl( {\frac {\mbox{\boldmath{$p$}}_i^2}{2m_i} } 
- \frac{Gm_0 m_i}{r_i} \biggr),  \nonumber \\
H_{*}  & = & \frac {\mbox{\boldmath{$p$}}_0^2}{2m_0} - \sum_{i=1}^n 
\biggl(  \frac{Gm_0 m_i}{r_{i0}} -  \frac{Gm_0 m_i}{r_i} \biggr), \nonumber \\
H_{int} & = &  - \sum _{ \genfrac{}{}{0pt}{}{i,j = 1}{i\ne j}}^n 
\biggl( \frac{Gm_i m_j}{r_{ij}} \biggr). \nonumber 
\end{eqnarray}
In the above equations, $G$ is the gravitational constant and 
$m_i$, $\mbox{\boldmath{$p$}}_i$, $r_i$ denote the mass, momentum and radial position
of the $i$-th object, where $i=0$ corresponds to the central star. 
$r_{ij}$ stands for the distance between the $i$-th and $j$-th objects.
$H_{Kep}$ is an integrable Hamiltonian corresponding to independent
Keplerian motions of three planets with respect to the barycenter.
$H_{*}$ is a correction originated from the orbital motion of the
central star itself by the attraction of the planets. $H_{int}$ expresses
the interactions between the planets.

For each model listed in Table~\ref{models}, we generate $10000$
different initial conditions, which have the same integrals of motion,
that is, the linear and angular momenta and total energy. This
constraint is important to explore the structure in the phase space.
The orbital phases of planets are given randomly. Since the radial location
of the innermost planet, $a_1$, is fixed, the remaining parameters, the
radial locations of the outer planets, $a_2$, $a_3$, and the velocity
of the central star, $ V_{sx}$, $V_{sy}$, are determined so that
the system should have the same values of the integrals of motion.
As a result of this constraint, the resultant initial radial locations
of the planets differ only slightly among $10000$ realizations.
 The planets have slight eccentricity (
 $e \sim 10^{-7}$ for $m_{pl}=10^{-7}$ and $e \sim 10^{-3}$ for $m_{pl}=10^{-3}$)
initially because 
of the non-zero velocity of the central star relative to the barycenter. 
Using the ensemble obtained in this way, we obtain various
distributions and take statistics thereof.
We summarize in Table~\ref{models} the input parameters ($a_1$ and the
integrals of motion) as well as the orbital separations and
eccentricities averaged over the $10000$ realizations for each model.

Numerical integrations are performed with the MERCURY6 package, which
was developed by \citet{mercury}. For the proto-planet system, we
integrate the orbital motions until the first close encounter occurs.
For arbitrarily chosen three models among $10000$ realizations, we 
continue the integration after the close encounter up to $10^7$ years 
in order to compute the Lyapunov exponent and power spectra in the 
post-encounter phase. For the Jupiter system, The integration is
terminated when one of the planets is ejected from the system. 

Before closing this section, we discuss the relative magnitude of 
each part of the Hamiltonian given in Eq.~(\ref{eq:hamiltonian})
 and the existence of KAM tori in our system.
The relative magnitude of $H_{*}$ to $H_{Kep}$ is always of the order
of $m_{pl}/M_{*}$ and so is the ratio of $H_{int}$ to $H_{Kep}$ unless
$r_{ij}/r_{i0}$ becomes as small as $\sim m_{pl}/M_{*}$. 
In our models, the value of $m_{pl}/M_{*}$ is $10^{-7}$ for the
proto-planet system or $10^{-3}$ for the Jupiter system and the
minimum value of $r_{ij}/r_{i0}$, which is achieved when two planets
have the same orbital phase, is $\sim (m_{pl}/M_{*})^{1/3} \gg m_{pl}/M_{*}$.
Thus, $H_{*}$ and $H_{int}$ are always small in our models.

As mentioned in Introduction, the KAM tori exist in the phase space of
the nearly integrable system, whose Hamiltonian is expressed as the sum of 
the integrable part $H_0(I)$ and the perturbation $\epsilon H(I,\theta)$: 
\begin{equation}
H(I,\theta  ) =H_0(I)+\epsilon H(I,\theta )\hspace{3mm}(\epsilon \ll 1),
\label{eq:near_integ}
\end{equation}
provided the perturbation is sufficiently small. Here $I$ and $\theta$
are the action and angle variable, respectively. 
It should be noted that our Hamiltonian does not
meet this condition. 
Hence we look for evidence
that this is really the case, employing the Lyapunov exponent, power
spectra of orbital elements and induction periods, 
in addition to the trajectories
in the  $I-\theta$ plane, 
which will be described in the following section.
 
%%%%%%%%%%%%%%%%%%%%%%%%%%%%%%%%%%%%%%%%%%%%%%%%%%%%%%%%%%%%%%%%%%%%%%%%%%%%%%%%%%%%%%%
\section{Analysis Methods}
%%%%%%%%%%%%%%%%%%%%%%%%%%%%%%%%%%%%%%%%%%%%%%%%%%%%%%%%%%%%%%%%%%%%%%%%%%%%%%%%%%%%%%%
In order to get some insight into the phase space structure of our
models, we employ three tools, that is, the Lyapunov exponent, the power
spectra of orbital elements and the distribution of the duration  
of the quasi-regular motions. The first two are useful to see the
degree of chaos of the system. If there remain KAM tori in the phase
space indeed, the system is expected to show both non-chaotic and
chaotic features, which will then be captured by these measures.
The last quantity will tell us if the planetary motions of our models 
can be interpreted as the induction phenomenon. In fact, the
distribution is expected to have a power law if it is really the case.

%%%%%%%%%%%%%%%%%%%%%%%%%%%%%%%%%%%%%%%%%%%%%%%%%%%%%%%%%%%%%%%%%%%%%%%%%%%%%%%%%%%
\subsection{Lyapunov exponent}
%%%%%%%%%%%%%%%%%%%%%%%%%%%%%%%%%%%%%%%%%%%%%%%%%%%%%%%%%%%%%%%%%%%%%%%%%%%%%%%%%%%
We compute the so-called maximum Lyapunov exponent. The maximum global
Lyapunov exponent, $\lambda_{global} $, is the local growth rate of the distance, 
$\parallel \delta {\bf x}(t)\parallel$, between adjacent orbits in the phase space and is
defined more precisely as
\begin{equation}
\lambda _{global} \equiv   \lim_{T\to\infty} \frac{1}{T}\log \frac{ \parallel \delta{\bf x}(T)\parallel}{\parallel \delta{\bf x}(0)\parallel }.
\label{eq:lypglo} 
\end{equation}
It is, of course, impossible in practice to compute the growth of the distance over the
infinite time. We calculate instead the following quantity as a function of
time and look into their behavior:
\begin{equation}
\lambda_{global}(t)  =  \frac{1}{t}\log \frac{ \parallel \delta{\bf x}(t)\parallel}{\parallel \delta{\bf x}(0)\parallel }.
\label{eq:lypglo_t} 
\end{equation}
The asymptotic limit of this function at $t \rightarrow \infty$ gives
the original Lyapunov exponent.

The integrable system has a null Lyapunov exponent while chaotic
systems have a positive Lyapunov exponent.
If the system is nearly integrable in particular, the Lyapunov
exponent defined by Eq.~(\ref{eq:lypglo_t}) oscillates around a small
but finite positive value and does not converge as 
$t \rightarrow \infty$.

In analyzing the system that shows both quasi-regular and chaotic
behaviors alternatively, it is useful to look also into the local Lyapunov
exponent, $\lambda_{local}$, which is the same local growth rate of the orbital separation in much shorter
times and is
expressed as  
\begin{equation}
\lambda _{local}(n,\tau ) \equiv \frac{1}{\tau}\log \frac{ \parallel \delta{\bf x}(n\tau)\parallel}{\parallel \delta{\bf x}((n-1)\tau)\parallel }, 
\end{equation} 
where $n$ specifies an interval with a period of $\tau$.
The choice of $\tau$ is rather arbitrary. It should be longer than the
typical orbital period but shorter than the time scale of the
quasi-regular or chaotic motions of interest. If chosen appropriately,
it will indicate the local degree of chaos.
% For computing these Lyapunov exponents, we used the algorithm
% developed by  \textit{et al. }(1985) 

In the following, the interval, $\tau$, for the local Lyapunov exponent
is chosen to be 100 years, which corresponds to $10 \sim 100$ times
the orbital periods. Both $\lambda_{global}$ and $\lambda_{local}$ are
obtained by numerically integrating the linearized equations of motion along 
the phase space orbit given by the integration of the equations of motion
 \citep{lyp}.

%%%%%%%%%%%%%%%%%%%%%%%%%%%%%%%%%%%%%%%%%%%%%%%%%%%%%%%%%%%%%%%%%%%%%%%%%%%%%%%%
\subsection{Power spectra of orbital elements}
%%%%%%%%%%%%%%%%%%%%%%%%%%%%%%%%%%%%%%%%%%%%%%%%%%%%%%%%%%%%%%%%%%%%%%%%%%%%%%%%%%%%%%
The power spectrum, $S(f)$, of an orbital element denoted by $z(t)$ 
is defined as
\begin{equation}
\label{eq:ps}
 S(f)\equiv \left|\int_{0}^{\infty} z(t)e^{if t /2\pi }dt
\right|^2
\end{equation}
and is also useful to characterize the chaotic system.
If one defines the autocorrelation function of $z(t)$ as
\begin{equation}
\label{eq:corr}
\Phi_z(\tau)\equiv \lim_{T\to\infty} \frac{1}{2T}\int_{-T}^{T} z(t)z(t+\tau)dt,
\end{equation}
its Fourier transform is equal to the power spectrum according to the 
Wiener-Khinchin theorem \citep[e.g.][]{w-k}: 
\begin{equation}
\label{eq:w-k}
 S(f)=\int_{-\infty}^{\infty} \Phi_z(\tau)e^{-if \tau /2\pi}d\tau.
\end{equation}
Hence by investigating the power spectrum, we can acquire the
knowledge of the temporal autocorrelation of the orbital element.

The integrable systems have a discrete spectrum with peaks
at the orbital periods provided appropriate variables are chosen. 
On the other hand, the chaotic systems have a rather featureless 
continuum spectrum. In particular, it is known that the nearly
integrable systems show in general a power-law spectrum in their low 
frequency regime as
\begin{equation}
\label{eq:ps_power}
S(f)\propto f^{-\nu} 
\end{equation}
indicating a long-time correlation for the variables.
 
In this paper, we discuss the power spectra of the orbital eccentricity, $e_i$,
for each planet in the system. We confirmed, however, that other
orbital elements such as the semi-major axis give a similar result.
% (\ref{}).

%%%%%%%%%%%%%%%%%%%%%%%%%%%%%%%%%%%%%%%%%%%%%%%%%%%%%%%%%%%%%%%%%%%%%%%%%%%%%%%%%%%%%%
\subsection{Distribution of induction periods \label{model-dist}}
%%%%%%%%%%%%%%%%%%%%%%%%%%%%%%%%%%%%%%%%%%%%%%%%%%%%%%%%%%%%%%%%%%%%%%%%%%%%%%%%%%%%%%
It is known for the induction phenomena that the distribution of the 
induction periods, $T$, or the duration of quasi-regular motions
generally obeys a power law, 
\begin{equation}
P(T)\propto T^{-\beta} \hspace{3mm}(\beta > 0), 
\label{eq:induction}
\end{equation}
for large $T$ \citep{bb90}. This implies that there is no
characteristic time scale for the induction period. 
This type of distribution can be obtained in the stagnant motion model 
by evoking a collection of tori, which have a self-similar 
distribution, in the so-called stagnant layer around a KAM torus in
the phase space \citep{aizawa}. Fig.~\ref{schematic1} illustrates
schematically the phase space structure assumed in the stagnant motion
model. The stagnant layer is filled with self-similarly distributed
tori, which trap the system around them for a long time. Since the
trapping time is scale free thanks to the self-similar distributions
of the tori, the power law is obtained for the induction period, which 
is nothing but the trapping time.   
 
We will use this feature to see if the planetary motions of our
concern are indeed the induction phenomenon. We expect that the
duration of quasi-regular motions corresponds to the induction period.
More precisely, we define the duration of the regular motion as the
interval from the start of the integration of motion until the first
close encounter between two planets. The close encounter means here that
the pair of planets has the distance between them smaller than their 
Hill radius. It is known from the previous papers and confirmed in
this paper that the transition from the regular motion to the chaotic
one occurs in general after the first close encounter \citep{cwb96}.
Incidentally, since the ejection of one planet occurs rather soon
after the encounter for the Jupiter system, we also study the distribution
of the time from the encounter to the ejection.

%%%%%%%%%%%%%%%%%%%%%%%%%%%%%%%%%%%%%%%%%%%%%%%%%%%%%%%%%%%%%%%%%%%%%%%%%%
 \section{Results}
%%%%%%%%%%%%%%%%%%%%%%%%%%%%%%%%%%%%%%%%%%%%%%%%%%%%%%%%%%%%%%%%%%%%%%%%%%

%%%%%%%%%%%%%%%%%%%%%%%%%%%%%%%%%%%%%%%%%%%%%%%%%%%%%%%%%%%%%%%%%%%%%%%%%%
\subsection{Lyapunov exponents}
\begin{figure}
  \begin{center}
    \begin{tabular}{c}
      \resizebox{140mm}{!}{\plotone{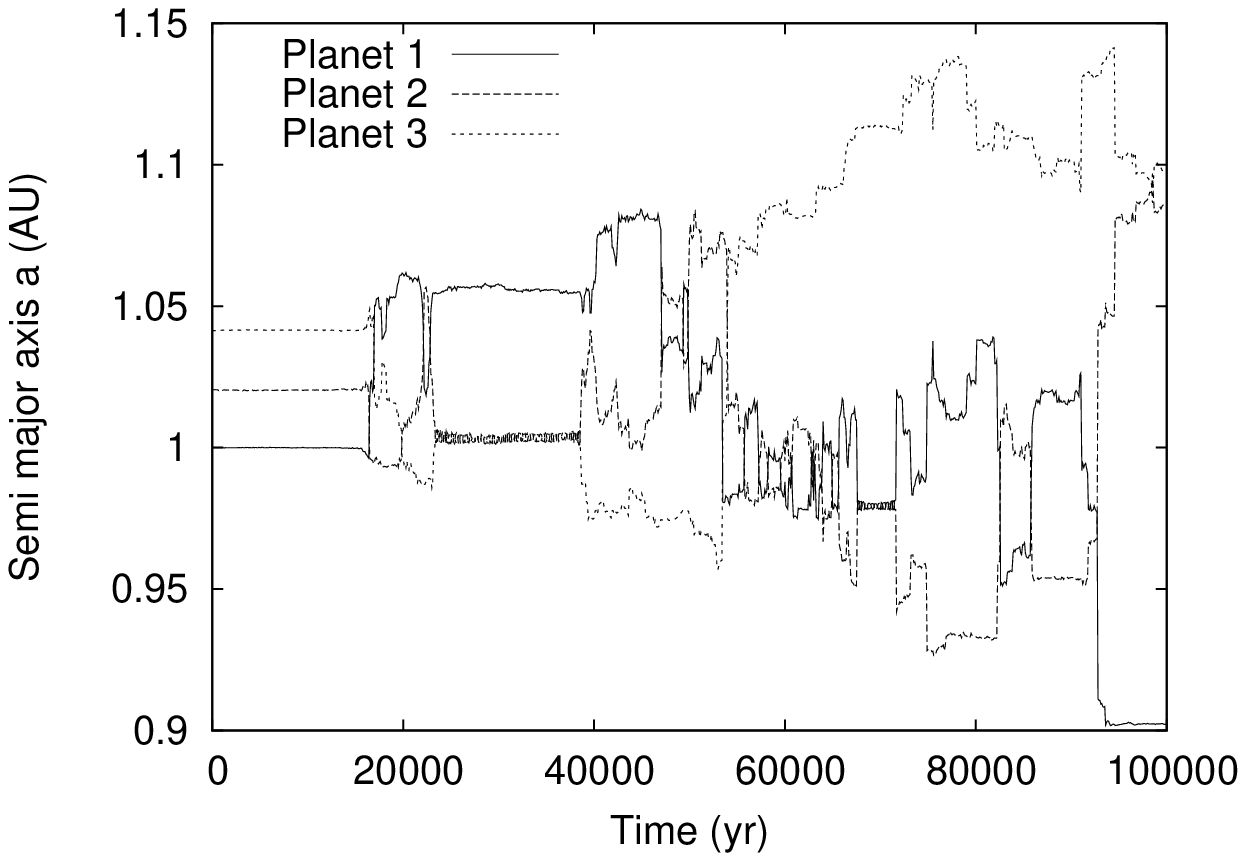}} \\
      \resizebox{140mm}{!}{\plotone{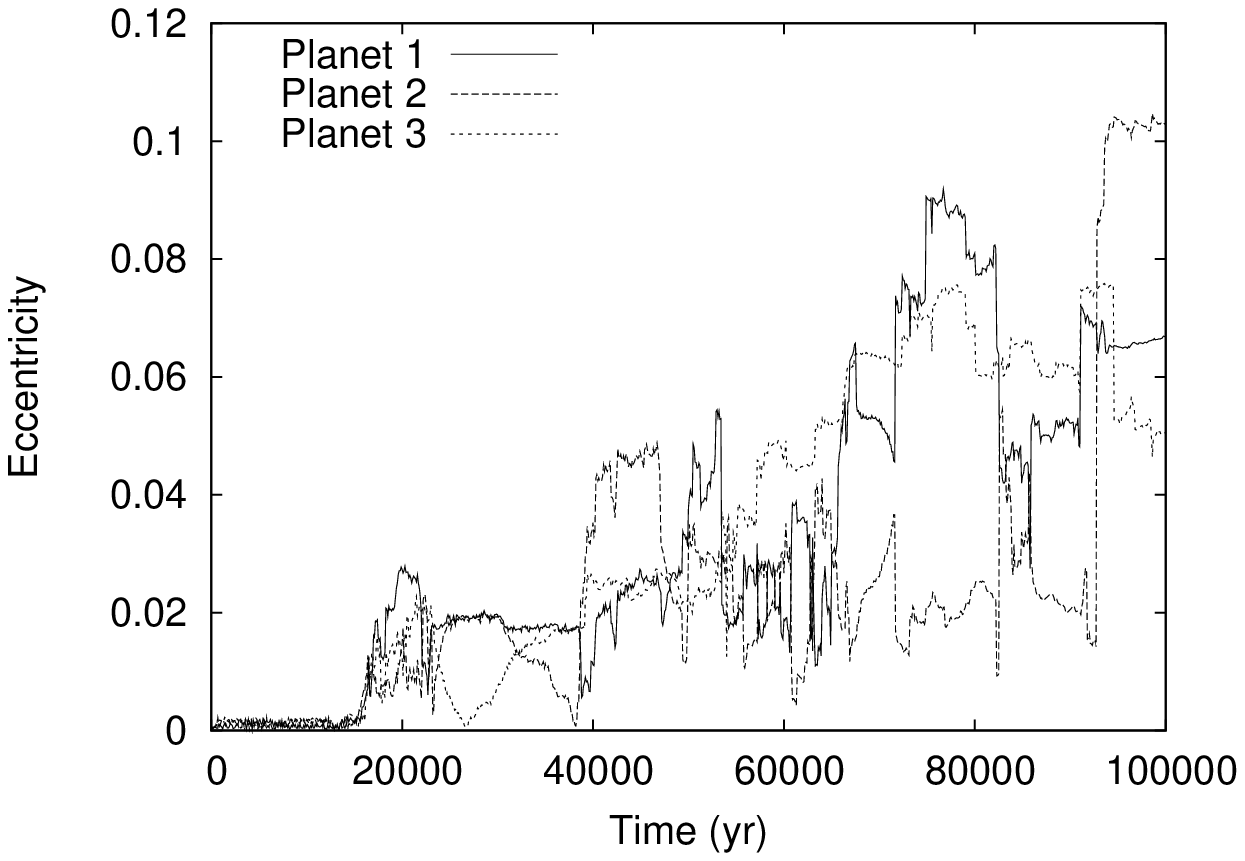}} \\
    \end{tabular} \caption{Time evolutions of the orbital semi-major axis 
(upper panel) and eccentricity (lower panel) of each planet 
for model $11$.
% with $m_{pl}=10^{-7}M_\odot$, $\Delta=5.0$.
}    \label{orbit_small}		
  \end{center}
\end{figure}

%%%%%%%%%%%%%%%%%%%%%%%%%%%%%%%%%%%%%%%%%%%%%%%%%%%%%%%%%%%%%%%%%%%%%%%%%%
We first show in Fig.~\ref{orbit_small} the evolution of the orbital
semi-major axis and eccentricity of each planet for 
model $11$, which is representative of the proto-planet system.
% with $m_{pl}=10^{-7} M_\odot$, $\Delta=5.0$ 
In this model, the
first close encounter between two planets occurs at $t = 16400$yr. 
Before the encounter, the planetary motions are almost regular and the 
orbital elements remain unchanged essentially. After the encounter,
on the other hand, they start to vary on short time scales. Then comes
a period ($t \simeq 23500-38500$yr) when the semi-major axises do not change very
much and the eccentricities for two planets vary rather
monotonically in this period. Thereafter the orbital elements change 
in time violently again. Note, however, that the quasi-regular phases,
although with much shorter periods, emerge intermittently.

\begin{figure}
  \begin{center}
    \begin{tabular}{c}
      \resizebox{140mm}{!}{\plotone{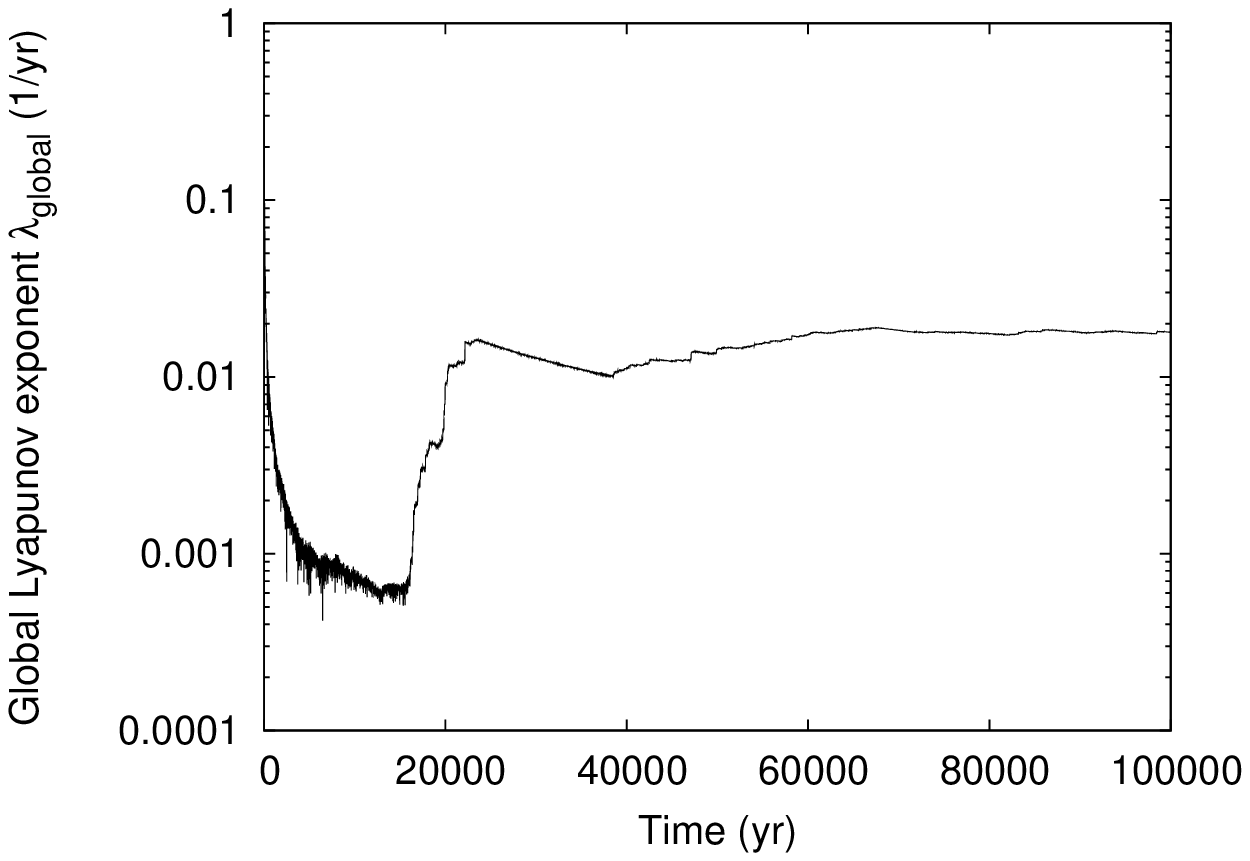}} \\
      \resizebox{140mm}{!}{\plotone{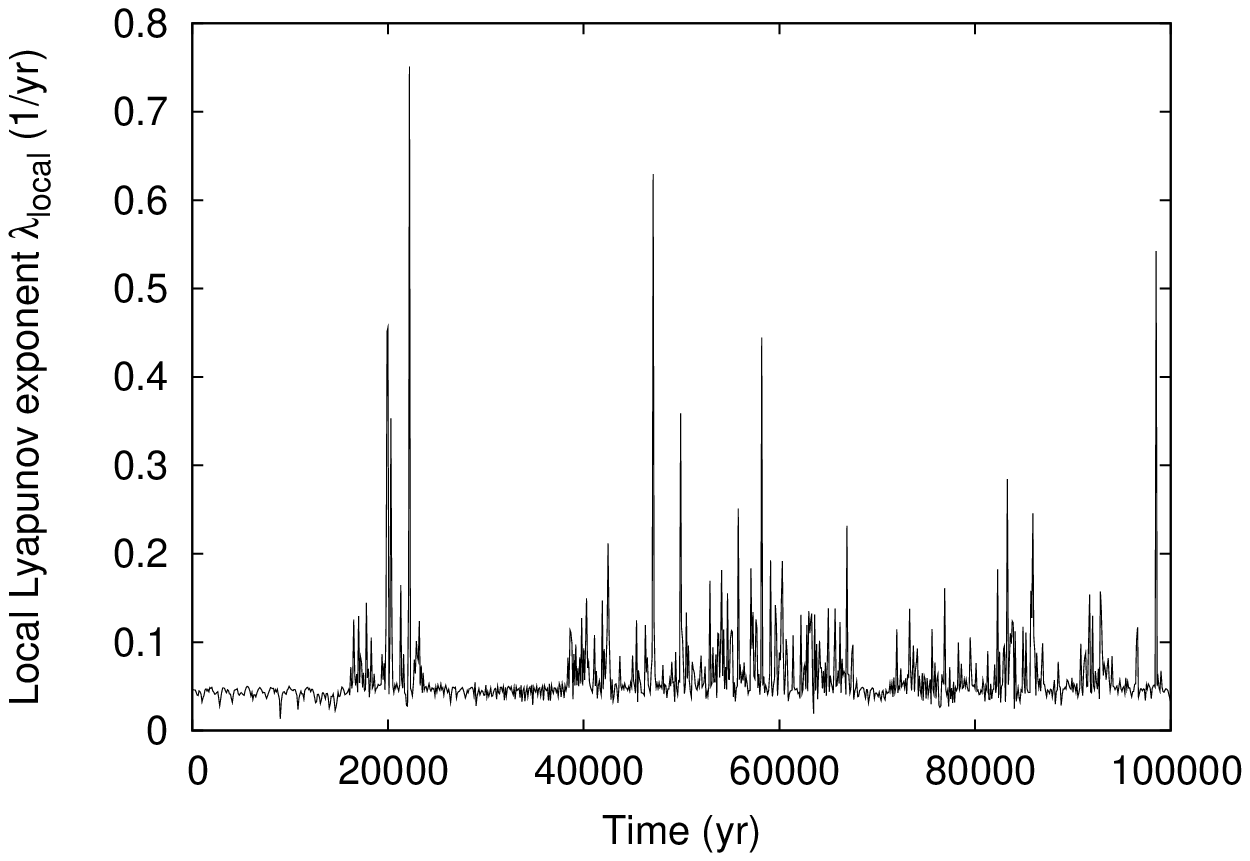}} \\
    \end{tabular}
\caption{Time evolutions of the global (upper panel) 
and Local (lower panel) Lyapunov exponents for the same model as in 
Fig.~\ref{orbit_small}. See the text for their definitions. 
}
    \label{lypsmall}		
  \end{center}
\end{figure}

In Fig.~\ref{lypsmall}, both the global and local Lyapunov exponents,
$\lambda_{global}(t)$ and $\lambda_{local}(n,\tau)$, are displayed as
a function of time for the same model as in Fig.~\ref{orbit_small}.
As mentioned earlier, the time interval $\tau$ for the integration of
the local Lyapunov exponent is set to be $100$yr, that is roughly 
100 times the orbital period for the proto-planet system, whose
innermost planet is initially located at 1AU. 
 It is clear that $\lambda_{global}(t)$ 
monotonically decreases toward zero before the close encounter 
($\lambda \simeq 0.0005$ just prior to the encounter), indicating that
the motions are non-chaotic (or very weakly chaotic) in this phase. 
After the encounter, on the other hand, $\lambda_{global}(t)$
increases drastically by about two orders of magnitude and fluctuates 
slowly around a constant value ($\sim 0.02$) thereafter. 
For comparison, we also employ MEGNO, another 
indicator of chaos suggested by \citet{cs00}, to estimate the global 
Lyapunov exponent. The values derived from MEGNO agree with 
the ones obtained above within $1\%$ typically.

The local Lyapunov exponent also shows that the orbital motions are
almost non-chaotic before the encounter. In fact, the almost constant
small value during this period is consistent with the evolution of the
global Lyapunov exponent for the first 100yr. It is also clear that the
local Lyapunov exponent shows remarkable peaks rather intermittently 
after the close encounter. Moreover, it is found by the comparison between
Figs.~\ref{orbit_small} and \ref{lypsmall} that $\lambda_{local}$
takes a small constant value when the semi-major axis of each planet
remains nearly constant in time. It is interesting to point out again that
the eccentricities of planetary orbits are not zero and change in time
during this period. This suggests that the planetary system is settled
to a quasi-stable configuration what is different from the initial
condition and has substantial orbital eccentricities. 

\begin{figure}
  \begin{center}
    \begin{tabular}{c}
      \resizebox{140mm}{!}{\plotone{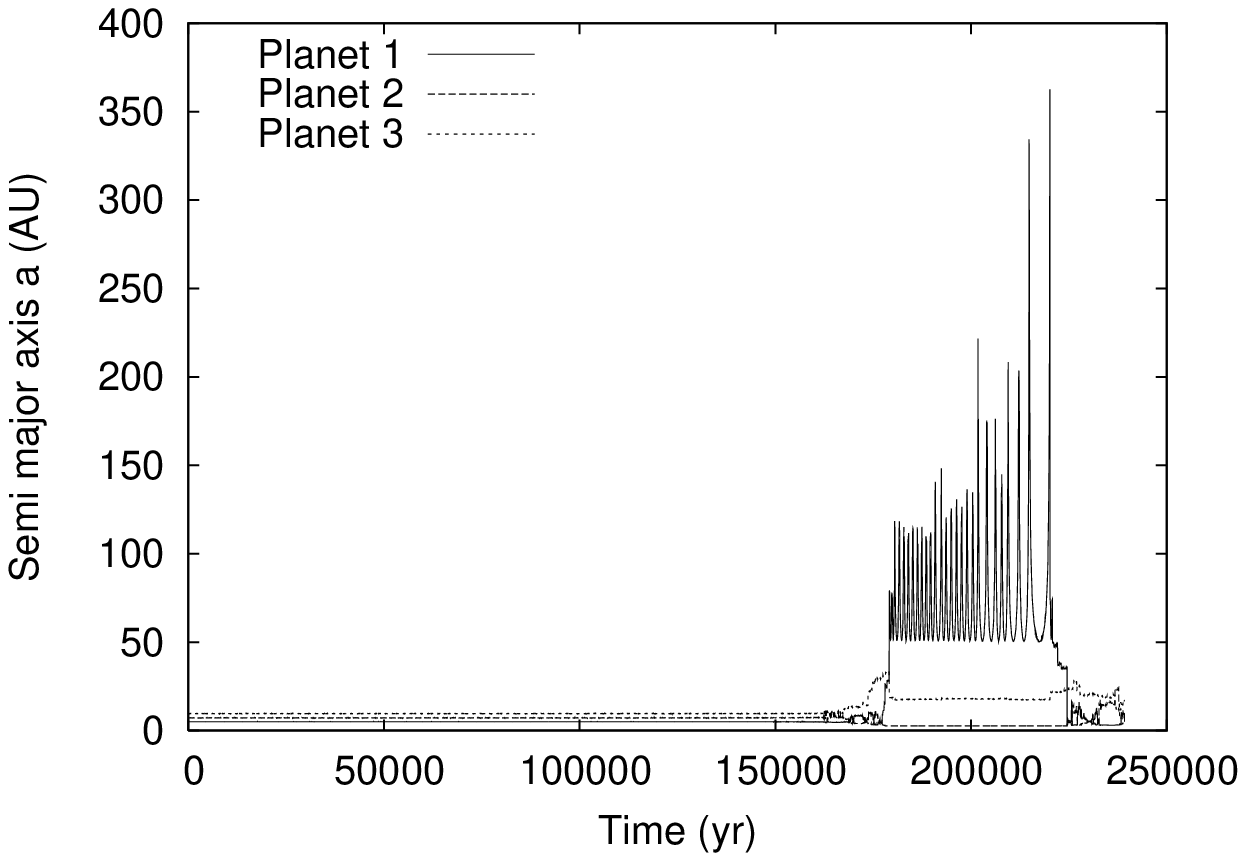}} \\
      \resizebox{140mm}{!}{\plotone{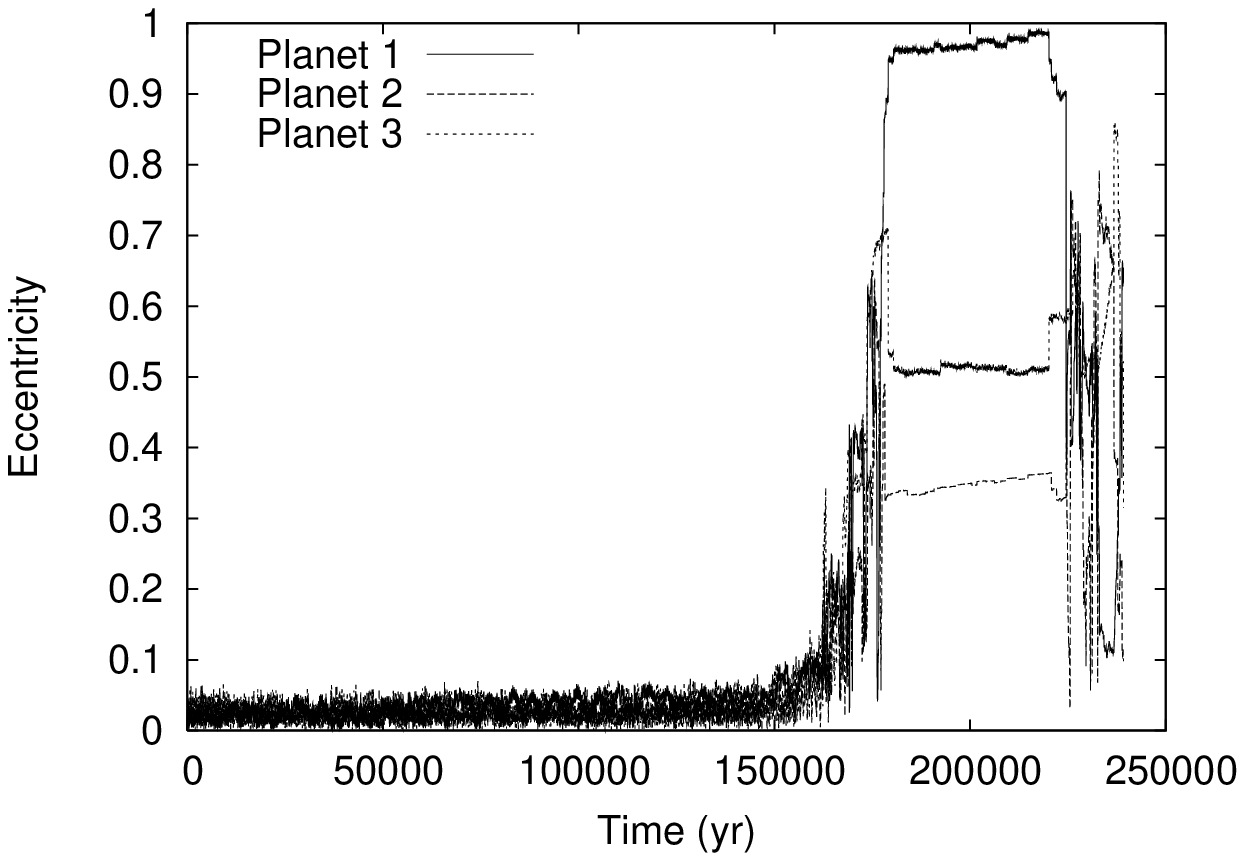}} \\
    \end{tabular} \caption{Time evolutions of the semi-major axis 
(upper panel) and eccentricity (lower panel) of each planet 
for model $17$.
% with $m_{pl}=10^{-3}M_\odot$, $a_1=5$, $a_2=7.25$, $a_3=9.5$AU.
}
   \label{orbit_big}		
  \end{center}
\end{figure}

In Fig.~\ref{orbit_big}, we show the orbital evolution of the Jupiter system
(model $17$). 
% with $m_{pl}=10^{-3}M_\odot$, $a_1=5$, $a_2=7.25$, $a_3=9.5$AU.
In this model, the first close encounter happens at $t=162416$yr and 
one of the planets is ejected from the system at $t=239279$yr.
Just as in the proto-planet system shown in Fig.~\ref{orbit_small}, the
semi-major axises remain almost unchanged in time and the
eccentricities oscillate around the initial value with small
amplitudes before the close encounter but they start to vary rapidly
in time after the encounter. 
It is noted that the amplitudes of
the variations are much larger for the Jupiter system than those for
the proto-planet system, the fact responsible for
the ejection of a planet in short times in the Jupiter system.

\begin{figure}
  \begin{center}
    \begin{tabular}{c}
      \resizebox{140mm}{!}{\plotone{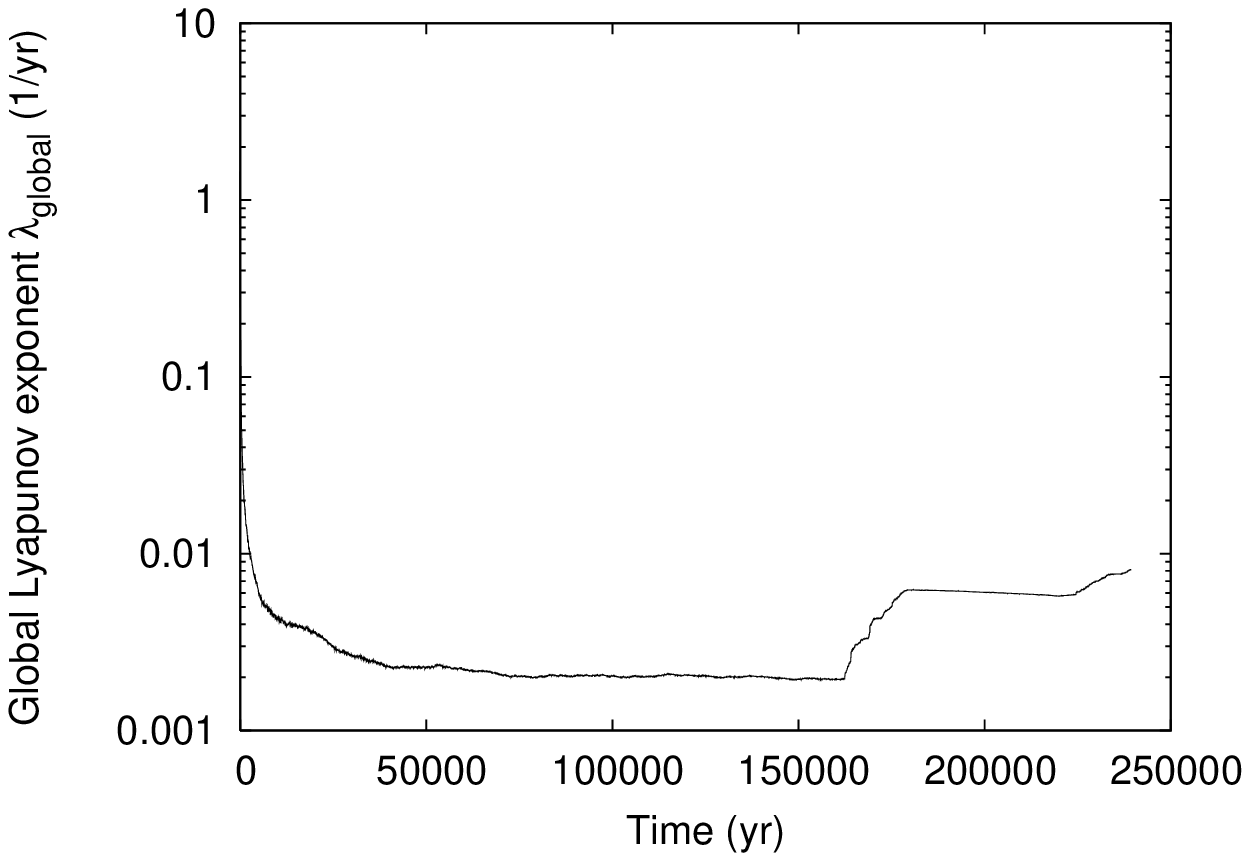}} \\
      \resizebox{140mm}{!}{\plotone{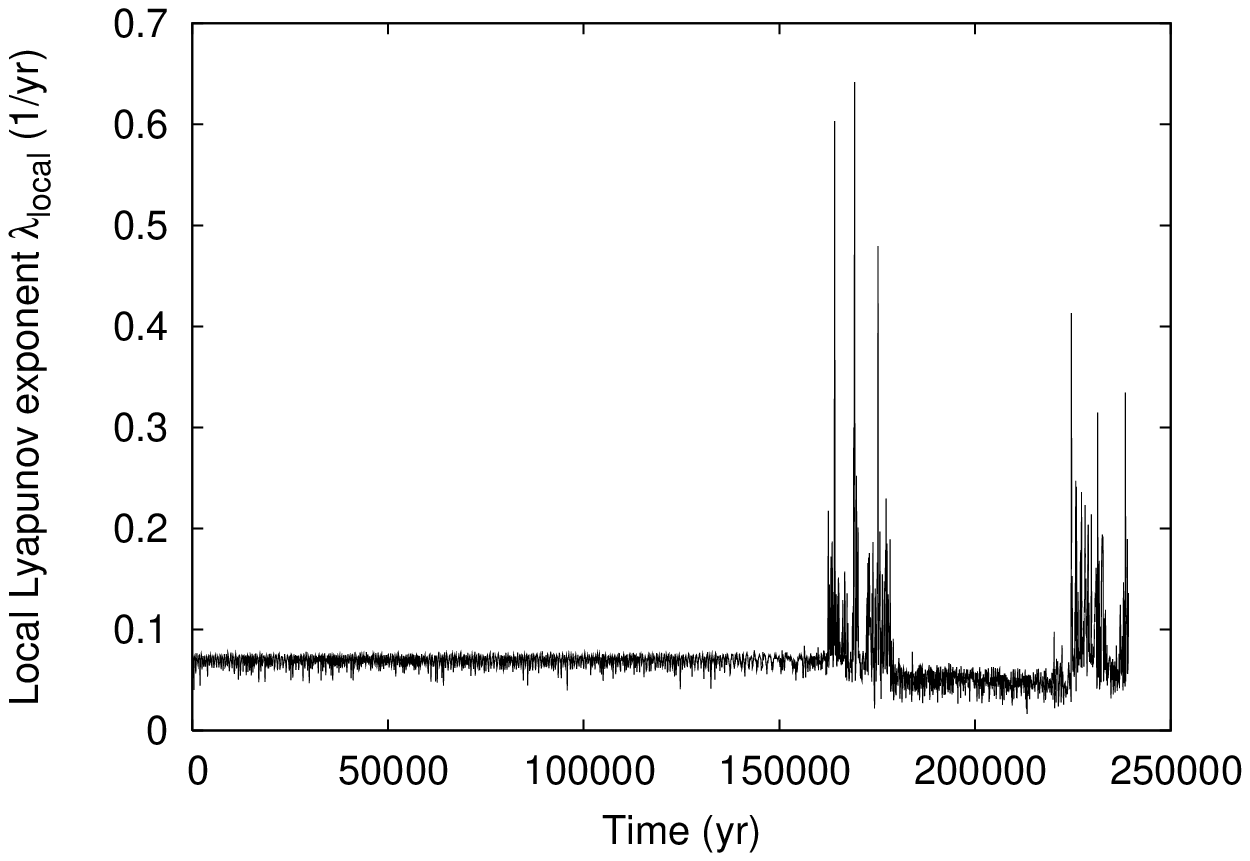}} \\
    \end{tabular}
\caption{Time evolutions of the global (upper panel) 
and local (lower panel) Lyapunov exponents for the same model as in 
Fig.~\ref{orbit_big}. 
}
    \label{lypbig}		
  \end{center}
\end{figure}

In Fig.~\ref{lypbig}, the global and local Lyapunov exponents are shown
for the same Jupiter system. Again we see the monotonic decrease of
the exponent for the first $\sim 20000$yr. In this case, however, the
exponent is then saturated and stays at a small but finite level.
This reflects the fact that the system is close to integrable but still 
non-integrable. The global Lyapunov exponent increases quickly after
the close encounter as in the proto-planet system. The local Lyapunov
exponent obtained for every 100yr, which is about 10 times the
orbital period for the Jupiter system, shows some intermittent spikes 
after the close encounter although the interval from the close
encounter to the ejection of a planet is rather short. It is also
confirmed that the violent variations of the orbital elements
occur in the spikes of the local Lyapunov exponent and that 
when $\lambda_{local}$ has a small, nearly constant value, the
semi-major axises are not changed very much whereas the eccentricities
are non-zero and fluctuate rather slowly. 
The Lyapunov 
exponent increases toward the ejection of the planet in this case.

The above-mentioned features in the orbital evolutions as well as in the
Lyapunov exponents are common to all the models. (The behavior of 
the Lyapunov exponent close to the ejection of a planet is an exception and
no clear trend can be seen). 
% To all the models, for which the Lyapunov exponents are obtained, the
% above mentioned features in the orbital evolutions as well as the
% Lyapunov exponents are common. 
This suggests that the underlying
structure in the phase space is not different very much from each
other. In particular, the quasi-regular motions before the close
encounter, which are very weakly chaotic at most, strongly suggest the
existence of the KAM torus. This may be true even of the periods that
occur intermittently after the close encounter, in which the local
Lyapunov exponent returns to a small value and the planetary motions 
become quasi-regular again. Then the following picture is inferred:
the phase space consists of a chaotic sea and KAM tori surrounded by
a stagnant layer that consists of satellite tori. The phase space 
orbits go from one system of KAM torus and stagnant layer to another 
through the chaotic sea. When the phase space orbit is moving around 
one of the KAM tori, the local Lyapunov exponent takes a small constant 
value whereas it becomes spiky once the phase space orbit moves into 
the chaotic sea.  

\begin{figure}
\begin{center}
\rotatebox{-90}{
 \resizebox{100mm}{!}{\plotone{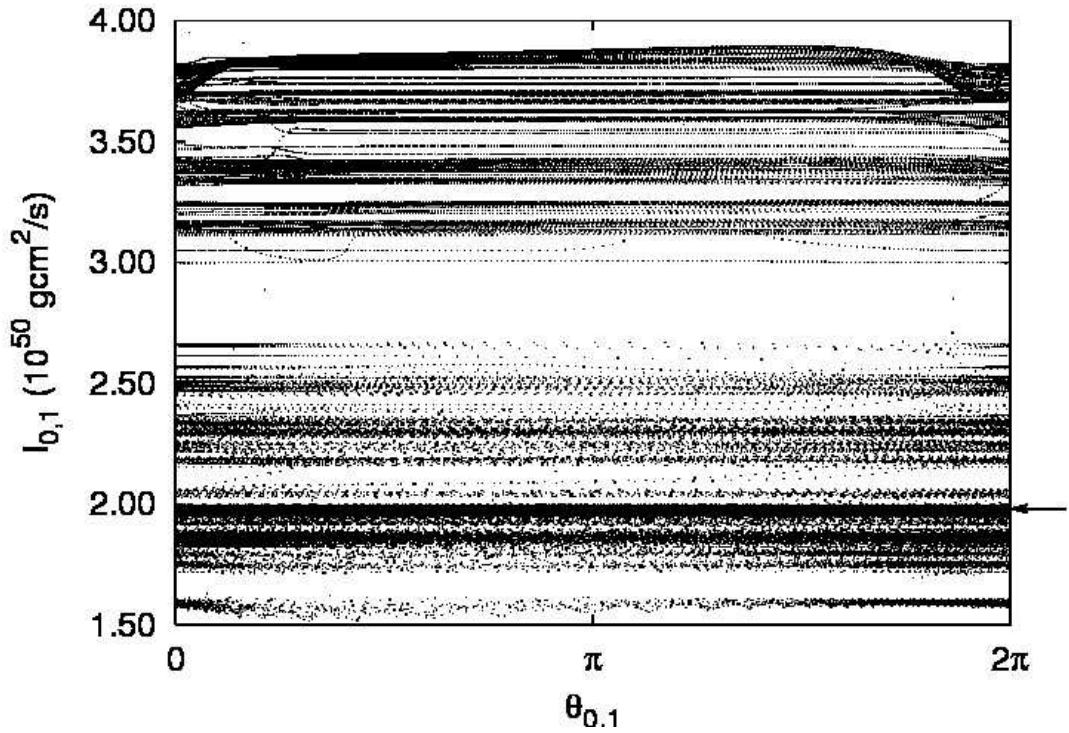}}}
\caption{The action ($I_{0,1}$) and angle ($\theta_{0,1}$) variables for the same model 
as in Fig.~\ref{orbit_big}. See Eq.~(\ref{eq:actionangle}) for the definitions of these variables. The horizontal line 
indicated by an arrow corresponds to the initial regular motion.}
    \label{actionangle}		
    \end{center}
\end{figure}

This picture is also supported 
by the plot in Fig.~\ref{actionangle} of the action ($I_0$) and angle ($\theta_0$) variables of the 
non-perturbed Hamiltonian $H_{Kep}$ in Eq.~(\ref{eq:hamiltonian}) .
These variables are given as 
\begin{eqnarray}\label{eq:actionangle}
I_{0,i} = m_i \sqrt{Gm_0 a_i},\\ 
\theta_{0,i} =  u_i - e_i \sin u_i.
\end{eqnarray}
In the above equations, the angle variable $\theta_{0,i}$ is called the mean anomaly and 
the so-called eccentric anomaly, $u_i$, is defined as 
\begin{equation}
\tan \frac{u_i}{2} = \sqrt{\frac{1-e_i}{1+e_i}}\tan{\frac{\phi _i}{2}},
\end{equation} 
where $e_i$ denotes the eccentricity and $\phi_i $ the orbital phase of the $i-$th planet measured in the barycentric coordinates.
In Fig.~\ref{actionangle}, we plot the action and angle variables for the innermost planet every ten steps.
The behavior of the variables for other planets is essentially the same.

One can recognize some horizontal bands and voids in the figure. These bands are regions, where the phase space orbit 
lingers, whereas the voids are passed through quickly. It is also seen that some of the bands undulate.
The densest band pointed by an arrow at $I_{0,1}=1.98 \times 10^{50}$ (gcm$^2$/s)) in the figure corresponds to the KAM torus 
of the initial regular motion and its satellite tori, whereas other bands represent other KAM tori and their satellites, which are visited 
by the phase space orbit during the evolution. The bands with undulation correspond to the motion approaching the ejection of 
a planet. Since the $I_{0,1}-\theta_{0,1}$ plane is filled by horizontal lines (or tori) everywhere uniformly if the perturbations are absent,
the voids can be interpreted as chaotic regions produced by the perturbations to $H_{Kep}$.

\begin{figure}
\begin{center}
\rotatebox{-90}{
 \resizebox{100mm}{!}{\plotone{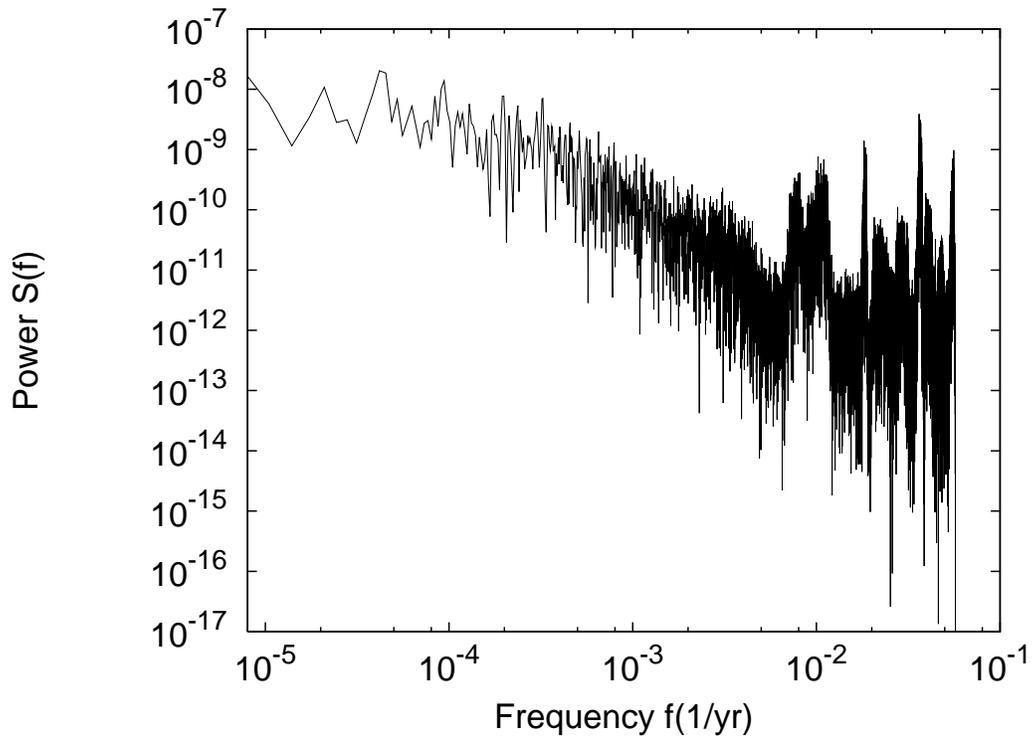}}}
\caption{The power spectrum of the action variable ($I_{0,1}(t)$) before the close encounter 
for the same model as in Fig.~\ref{orbit_big} .
}
    \label{spaction}		
    \end{center}
\end{figure}

Looking more closely, one finds that all the lines composing a band that corresponds to a quasi-regular motion are oscillating with 
finite amplitudes. These oscillations are studied by the Fourier analysis and the power spectrum for the initial quasi-regular motion
is plotted in Fig.~\ref{spaction}. It is found that the power spectrum obeys a power-law over a wide frequency range. This means that there is no
characteristic frequency scale, the fact which seems to be consistent with the stagnant motion model, in which satellite tori are supposed to 
exist around a KAM torus with a fractal size distribution.

In the following we will further look for evidence
that the dynamics in these periods, that is, a relatively long
quasi-regular motions followed by an abrupt transition to chaotic
motions can be interpreted as an induction phenomenon indeed.

%%%%%%%%%%%%%%%%%%%%%%%%%%%%%%%%%%%%%%%%%%%%%%%%%%%%%%%%%%%%%%%%%%%%%%%%%%%%%%%%%%%%%%%%
\subsection{Power spectra of orbital eccentricities}
%%%%%%%%%%%%%%%%%%%%%%%%%%%%%%%%%%%%%%%%%%%%%%%%%%%%%%%%%%%%%%%%%%%%%%%%%%%%%%%%%%%%%%%%

\begin{figure}
   \begin{center}
    \begin{tabular}{c}
           \resizebox{140mm}{!}{\plotone{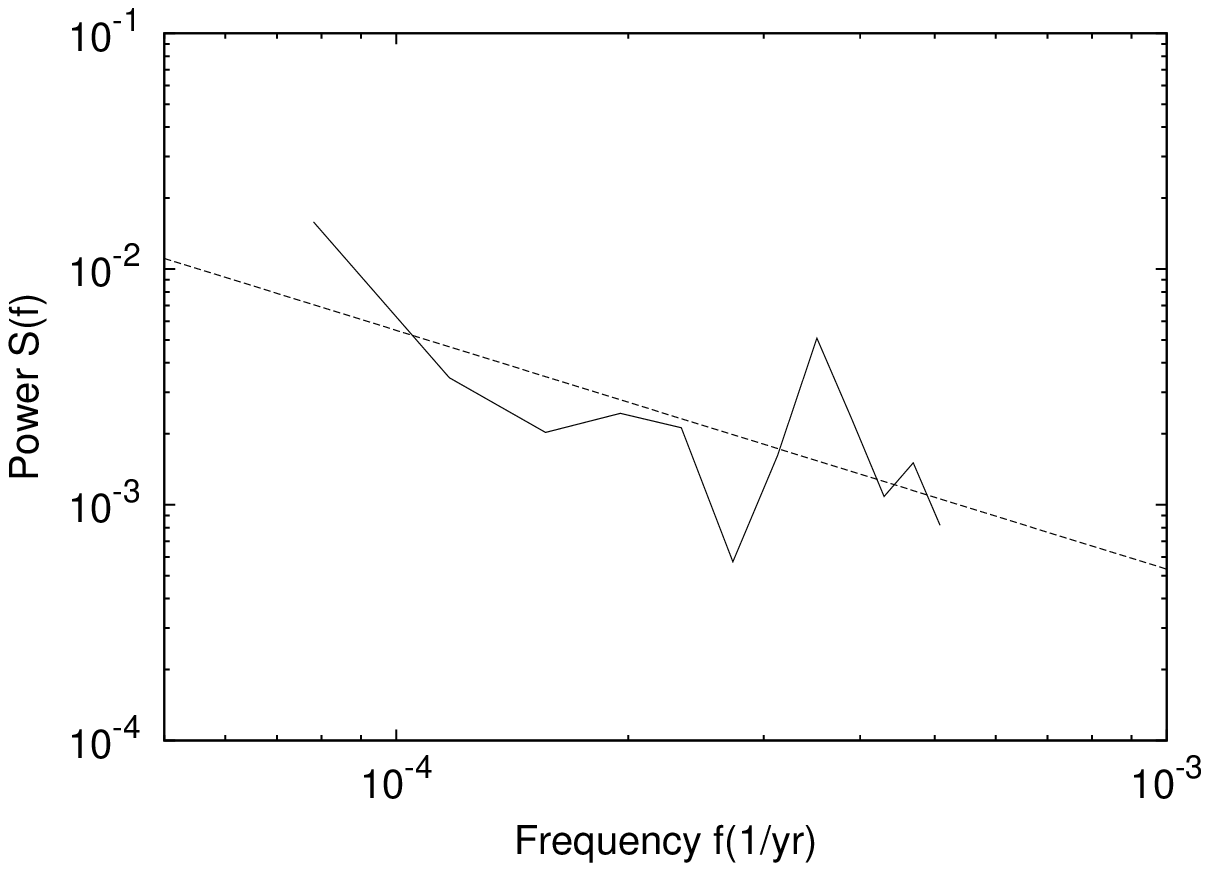}} \\
           \resizebox{140mm}{!}{\plotone{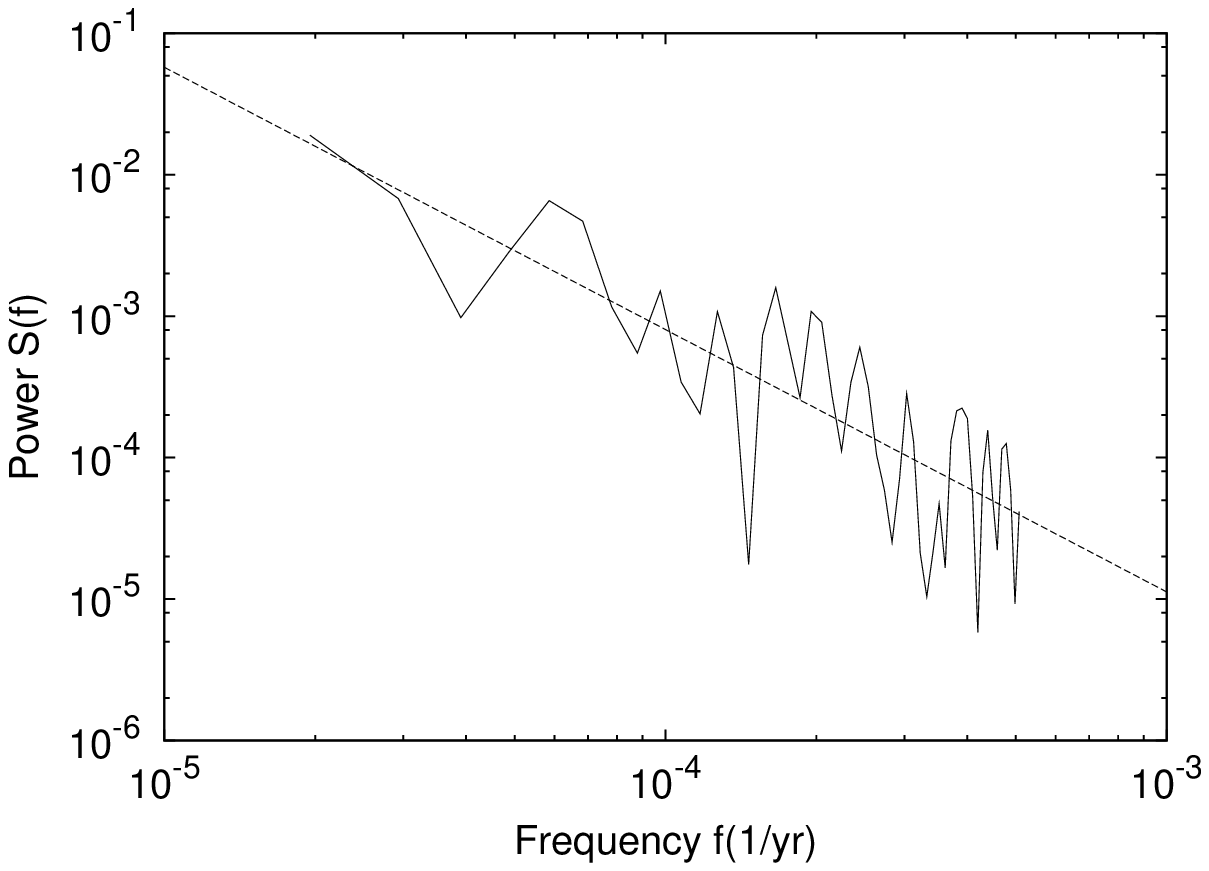}} \\
    \end{tabular}
\caption{ Power spectra $S(f)$ of the orbital
    eccentricity of the innermost planet before (upper
    panel) and after (lower panel) the close encounter for model $11$. 
 The straight
 lines are the fit to the data. The power-law indices are 1.01 and
 1.85 before and after the encounter, respectively.
}
    \label{sp_ex_small}		
   \end{center}
\end{figure}

In Fig.~\ref{sp_ex_small}, we show the power spectra of the orbital
eccentricity of the innermost planet before and after the close
encounter for model $11$. This is a representative model for the
proto-planet system and the one we employed to demonstrate the
behavior of the Lyapunov exponents in Fig.~\ref{lypsmall}. This model
also has a merit that it has a relatively long duration up to the
first close encounter. Although the data are quite noisy, they can be
roughly fit by the power law, $\propto 1/f^{\nu}$. It is apparent that
the power-law indices are different between the two phases. We obtain 
$\nu = 1.01$ before the encounter whereas the spectrum becomes
steeper with $\nu = 1.85$ thereafter.

\begin{figure}
   \begin{center}
    \begin{tabular}{c}
           \resizebox{140mm}{!}{\plotone{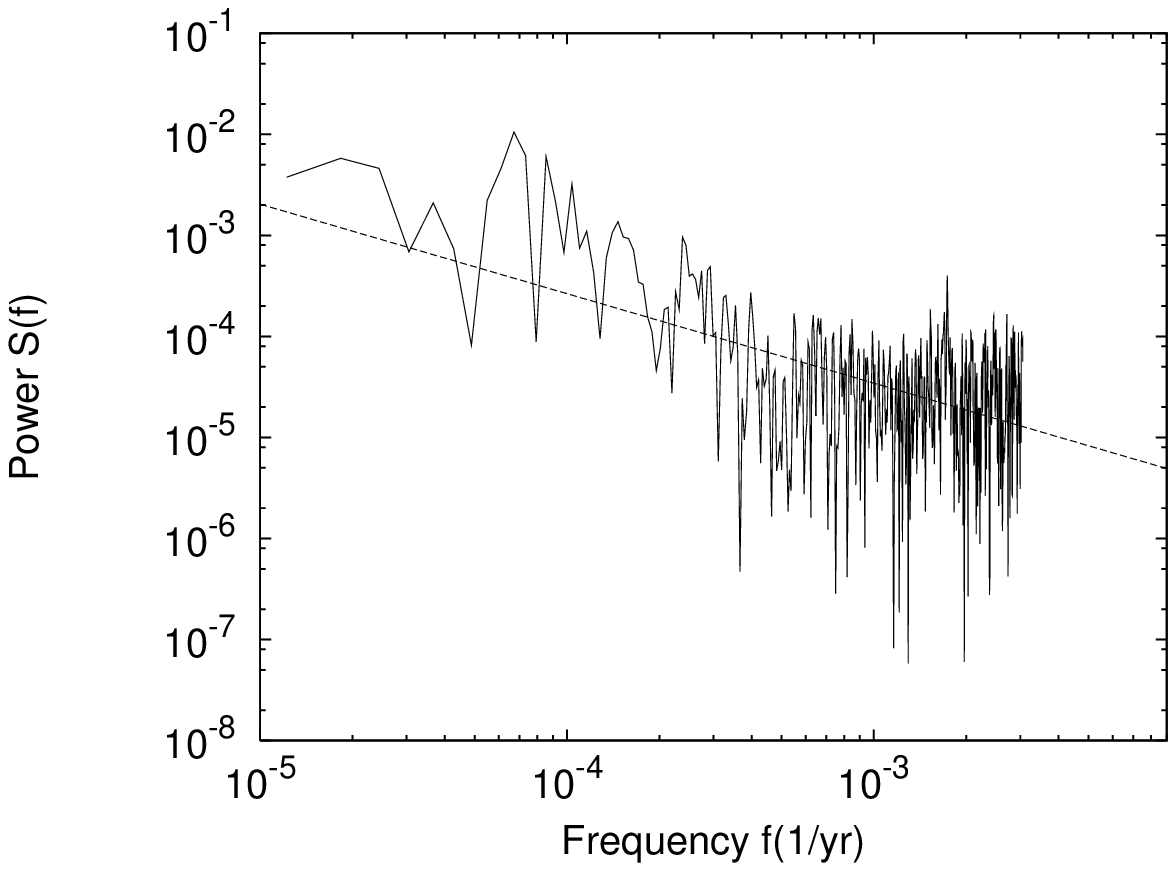}} \\
           \resizebox{140mm}{!}{\plotone{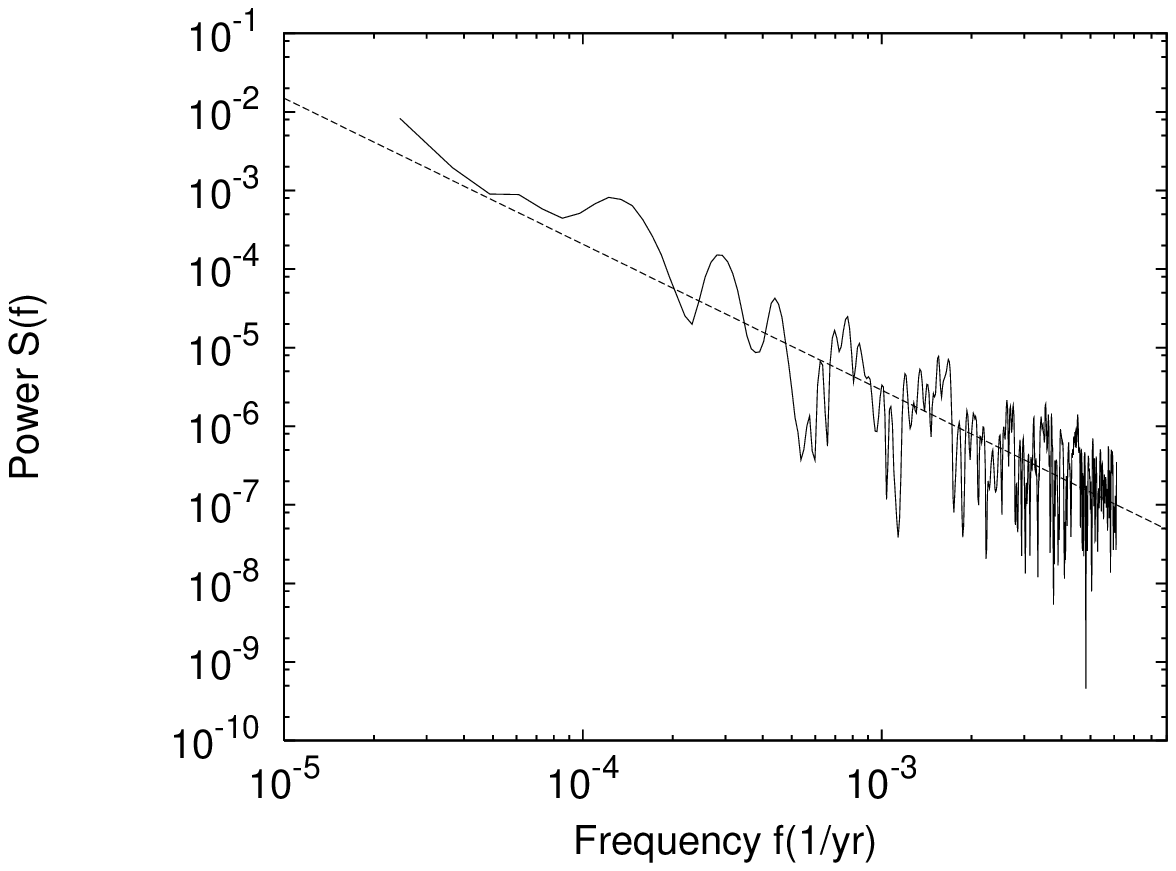}} \\
    \end{tabular}
\caption{ Power spectra $S(f)$ of the orbital
    eccentricity of the innermost planet before (upper
    panel) and after (lower panel) the close encounter for model $17$. 
 The straight lines are the fit to the data. The power-law indices are 0.88 and
 1.85 before and after the encounter, respectively.
}
    \label{sp_ex_big}		
   \end{center}
\end{figure}

The same trend can be seen also for the Jupiter system, whose typical 
results are displayed in Fig.~\ref{sp_ex_big} for model $17$.
Again the spectra are fit by the power law approximately both before
and after the encounter and the spectral indices are $\nu = 0.88$
before the encounter and $\nu = 1.85$ thereafter. 
% In general, the 
% orbital numbers up to the encounter is larger for the Jupiter system
% than for the proto-planet system. 
% This is the reason why the spectrum
% given in the upper panel of Fig.~\ref{sp_ex_big} has finer features
% than that in the upper panel of Fig.~\ref{sp_ex_small}. 
% Note also that the
% post-encounter phase was computed up to the ejection of a planet in
% this case. 
The spectrum
given in the upper panel of Fig.~\ref{sp_ex_big} has finer features
than that in the upper panel of Fig.~\ref{sp_ex_small} because 
pre-encounter phase is longer for the model in Fig.~\ref{sp_ex_small}
than for that in Fig.~\ref{sp_ex_big}. Note also that the
post-encounter phase was computed up to the ejection of a planet in
this case. 

In Tables~\ref{spsmall} and \ref{spbig}, we summarize the power-law spectral
indices of the arbitrarily chosen three realizations for each
model both before the encounter (1st phase in the table) and after the
encounter (2nd phase). Although we again employ the orbital eccentricity
of the innermost planet, the orbits of outer planets behave
similarly. Table~\ref{spsmall} gives the results for the proto-planet
system while Table~\ref{spbig} corresponds to the Jupiter system. For some
of the models, the duration of the phase is too short to obtain the
spectral index and "-" is put instead of the spectral index for
them. Although the spectral indices vary substantially even among different
realizations for the same model, it is clear that phase 2 has
spectral indices clustered around $\nu = 2$ whereas phase 1 has
smaller indices, $ 0 \lesssim  \nu \lesssim  1.69 $ , in general. The 
average spectral indices over all the models in Table~\ref{spsmall}
are 0.84 and 1.84 for the phases 1 and 2, respectively. The counter
parts for the Jupiter system given in Table~\ref{spbig} are $0.91$
before the encounter and $1.77$ after the encounter.

The power-law spectrum is a characteristic feature of the 
stagnant motions although the power-law indices are not specified by the theory. 
The time variations of the orbital elements are induced by the energy
exchange among planets. 
The power-law $\propto 1/f^2$ observed for the
post-encounter phase indicates that the time evolution of the orbital 
eccentricity is a Brownian motion with the root mean square being 
$\propto t^{1/2}$. 
On the other hand, the pre-encounter phase has a
smaller spectral index in general. 
Although the numbers of the planets in the system are different, our
results on the growth rate of eccentricity in the post-encounter phase
are consistent with \citet{zls07}, who studied the system with 50
planets and claimed that the eccentricity of planetary orbits roughly 
evolves as $\langle e^{2} \rangle ^{1/2} \propto t^{1/2}$ both before and after the
encounter. As mentioned above, however, we found for our system with
three planets different power-law indices before the
encounter. 
The smaller power-law indices of $e(t)$ in the pre-encounter phase in our models 
might correspond to the so-called fractional Brownian motion, for which the root mean square 
$\langle e^{2} \rangle ^{1/2}$ grows more slowly than for the ordinary Brownian motion \citep{mv68}.

%%%%%%%%%%%%%%%%%%%%%%%%%%%%%%%%%%%%%%%%%%%%%%%%%%%%%%%%%%%%%%%%%%%%%%%%%%%%%%%%%%%%%%%%
\subsection{Distribution of induction periods}
%%%%%%%%%%%%%%%%%%%%%%%%%%%%%%%%%%%%%%%%%%%%%%%%%%%%%%%%%%%%%%%%%%%%%%%%%%%%%%%%%%%%%%%%

%%%%%%%%%%%%%%%%%%%%%%%%%%%%%%%%%%%%%%%%%%%%%%%%%%%%%%%%%%%%%%%%%%%%%%%%%%%%%%%%%%%%%%%%
\subsubsection{proto-planet system}
%%%%%%%%%%%%%%%%%%%%%%%%%%%%%%%%%%%%%%%%%%%%%%%%%%%%%%%%%%%%%%%%%%%%%%%%%%%%%%%%%%%%%%%%

\begin{figure}
\begin{center}
    \begin{tabular}{ll}
      \plottwo{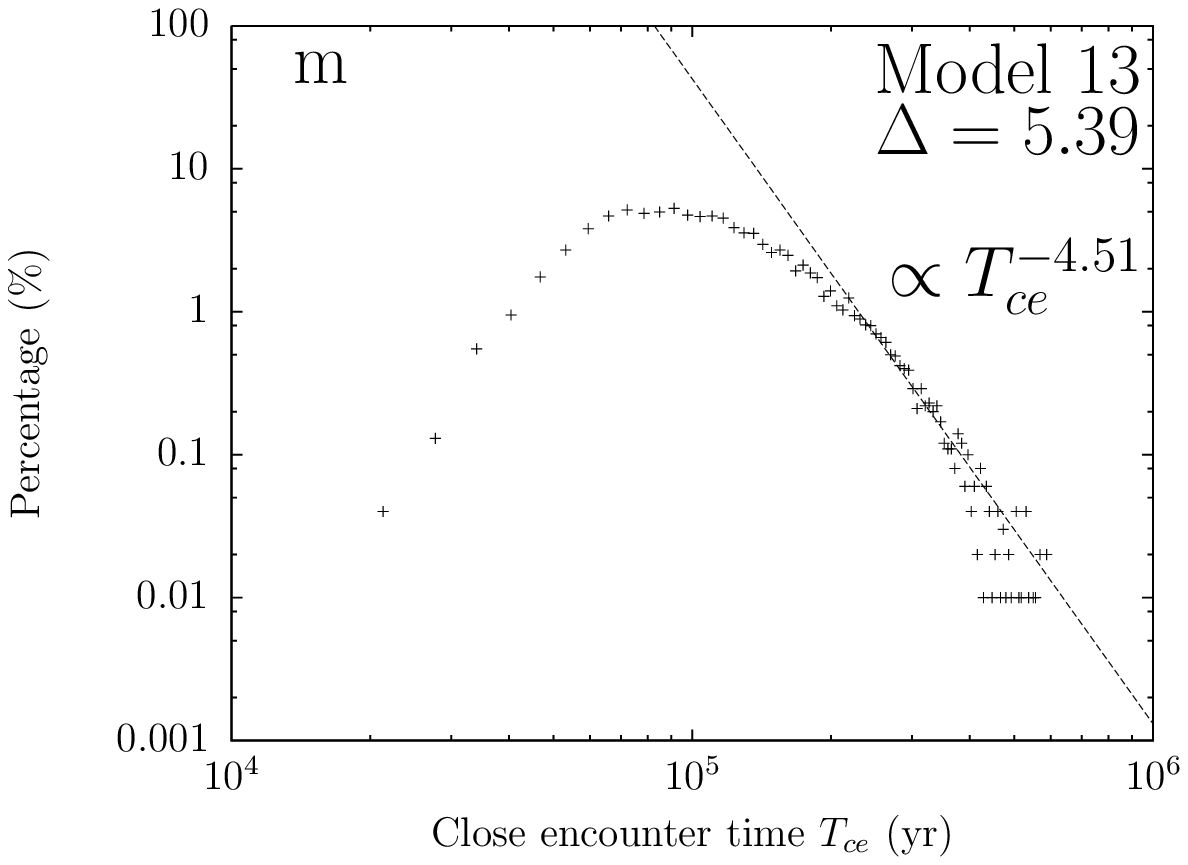}{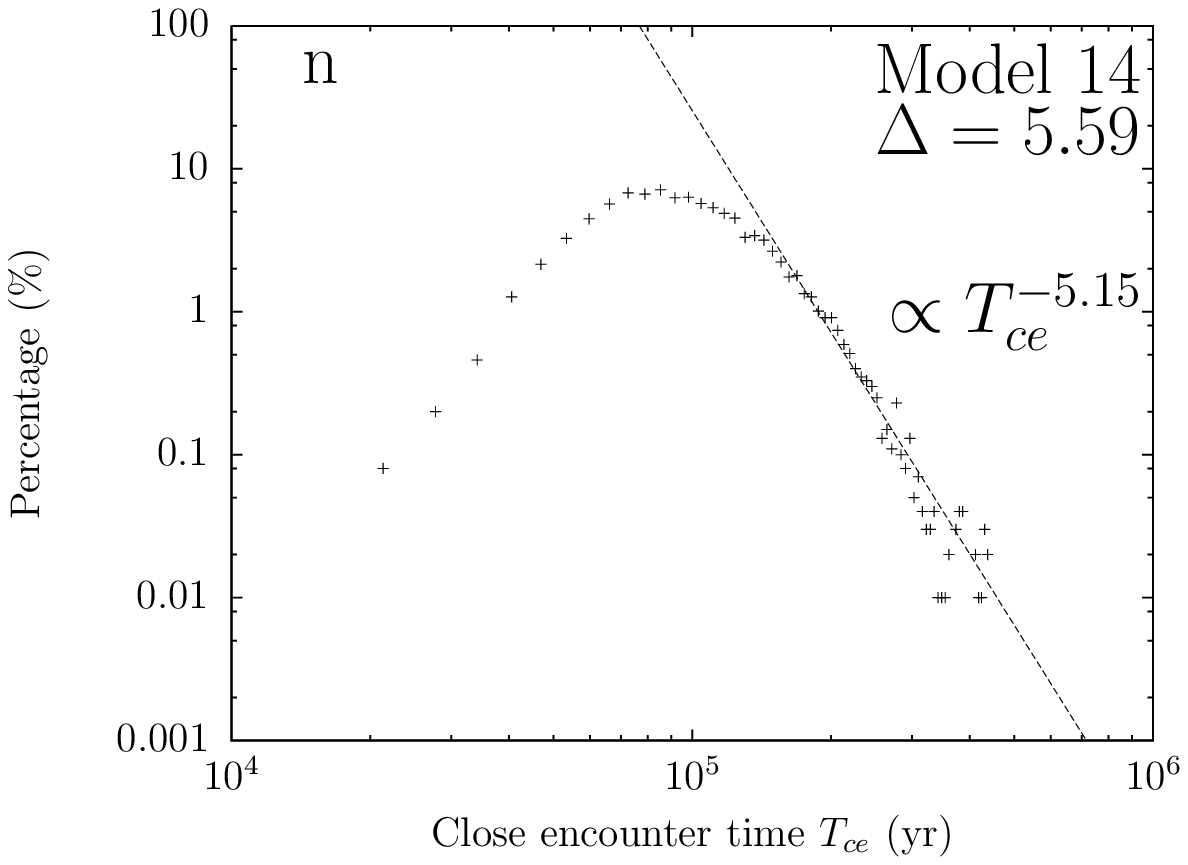} \\
      \plottwo{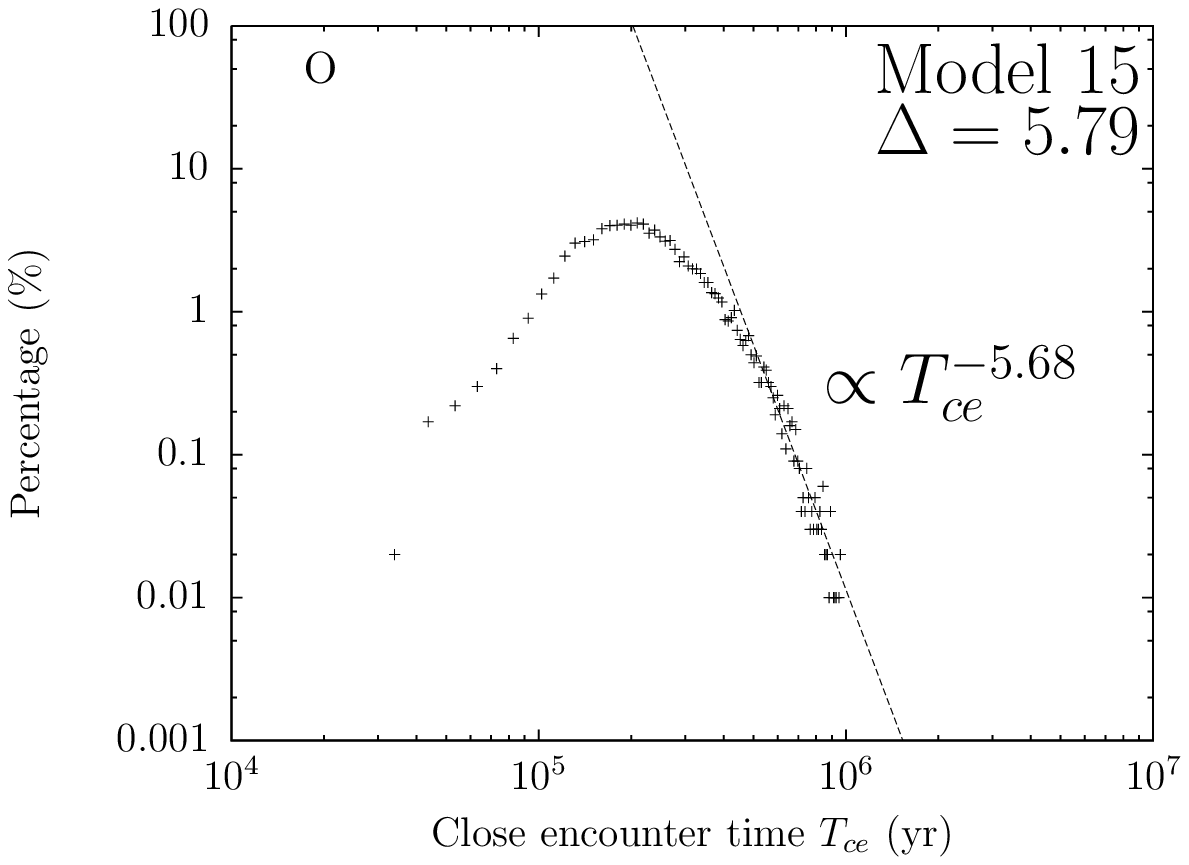}{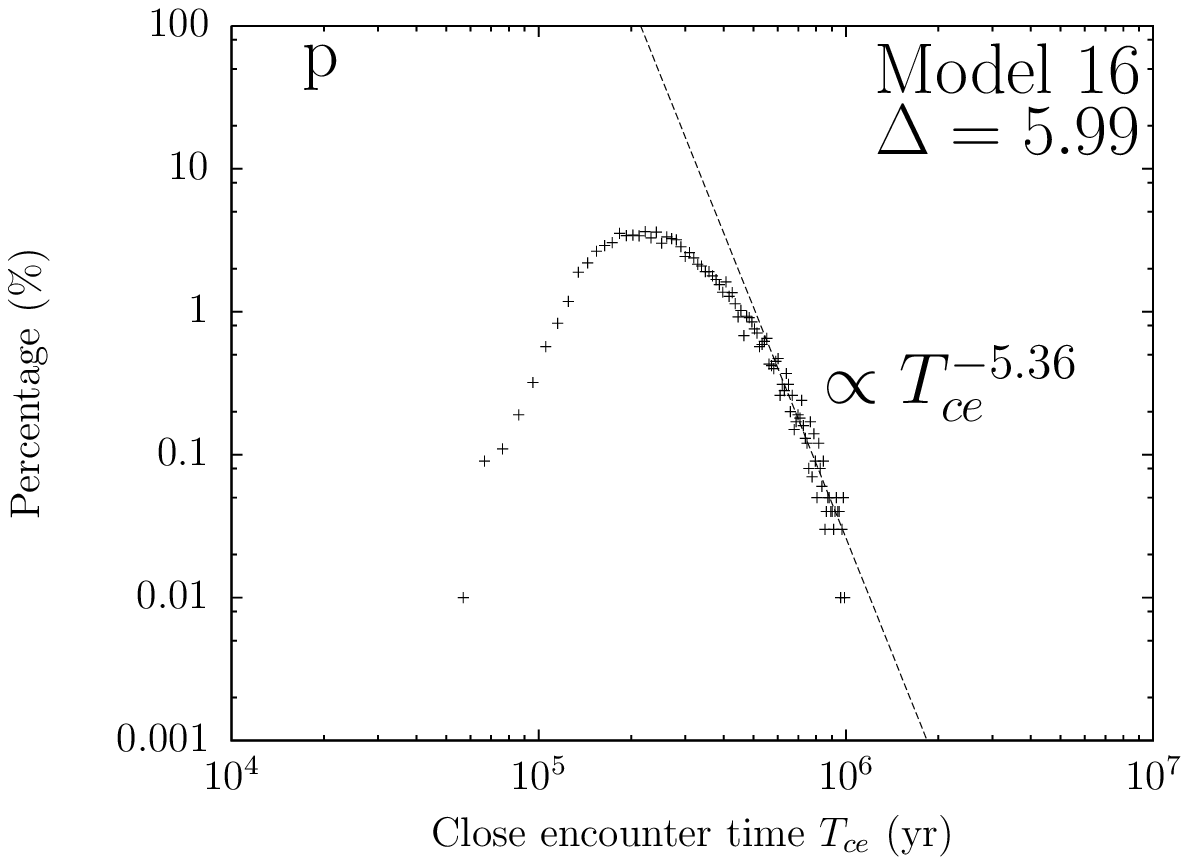} \\
    \end{tabular}
    \caption{Normalized distributions of the duration of the
      pre-encounter phase, $T_{ce}$, for models~1 to 16. 
      The straight lines are the fit by power law to the data in the 
      long-duration regime.}
    \label{distsmall}
\end{center}
\end{figure}

As mentioned in section~\ref{model-dist}, the distributions of the
durations of various phases should be one of the key
ingredients if the dynamics were to be interpreted as an induction
phenomenon and described by the stagnant motion model. For the
proto-planet systems we study the statistics of the duration,
$T_{ce}$, of the pre-encounter phase, where the orbits are nearly circular.
We show in Fig.~\ref{distsmall} the distribution of $T_{ce}$ for
$10000$ realizations of each model for the proto-planet systems listed
in Table~\ref{models}. For the plots we employed $100$ equal bins
between the maximum and minimum values of $T_{ce}$. For some models with 
small initial orbital separations, these two values are quite
different and, as a result, there are some bins with very small
populations near the longest $T_{ce}$.

Except for the model with the smallest $\Delta$, the distribution has
a peak, which shifts to longer times as $\Delta$ becomes larger. 
It is also clear from the log-log plot that the dispersion around the
peak gets larger, too, as $\Delta$ becomes greater. We are particularly
interested in the long time regime, where the stagnant motion model
predicts power laws, which implies that there is no characteristic
time scale for the trapping in the stagnant layer and is supposed to
be a consequence of the self-similar distribution of tori in the
stagnant layer. In the figure, the straight lines are the power-law
fit to the long-time part of the distributions. We employ 30 data
points down from the one with the longest $T_{ce}$. The bins with 
$T_{ce}$ shorter than that at the peak or those with the population 
of less than 2\% are discarded. If there are less than 30 data points
that satisfy the criteria, all of them are used. 
% Although the stagnant
% motion model does not predict a particular value for the power-law
% index, we give the number for each panel, which lies between $-5.68$ 
% to $-2.27$.
The obtained power-law indices lie between $-5.68$ to $-2.27$ as shown
in each panel, which are expected to reflect the difference in 
the phase space structures.
  
It is obvious that the distributions are deviated from power laws both
at the short and long durations. The initial conditions prepared so
that planets are initially in regular motions may lead to the
underestimation of the duration time, since it might have cut short
the earlier portion of the pre-encounter phase. Note, however, that there
is no reason that we expect power laws for short time scales. As for
the longer time scales, on the other hand, power laws are expected if
the stagnant motion model can be applied. We suspect that the main
reason for the deviation from the power law for the very long
durations is that the number of realizations, that is $10000$ for each
model, are not large enough. In fact, only a small number of
realizations are contributing to the longest time
portion of the distributions, in which the deviation from the power
law is remarkable. 

\begin{figure}
\begin{center}
\plotone{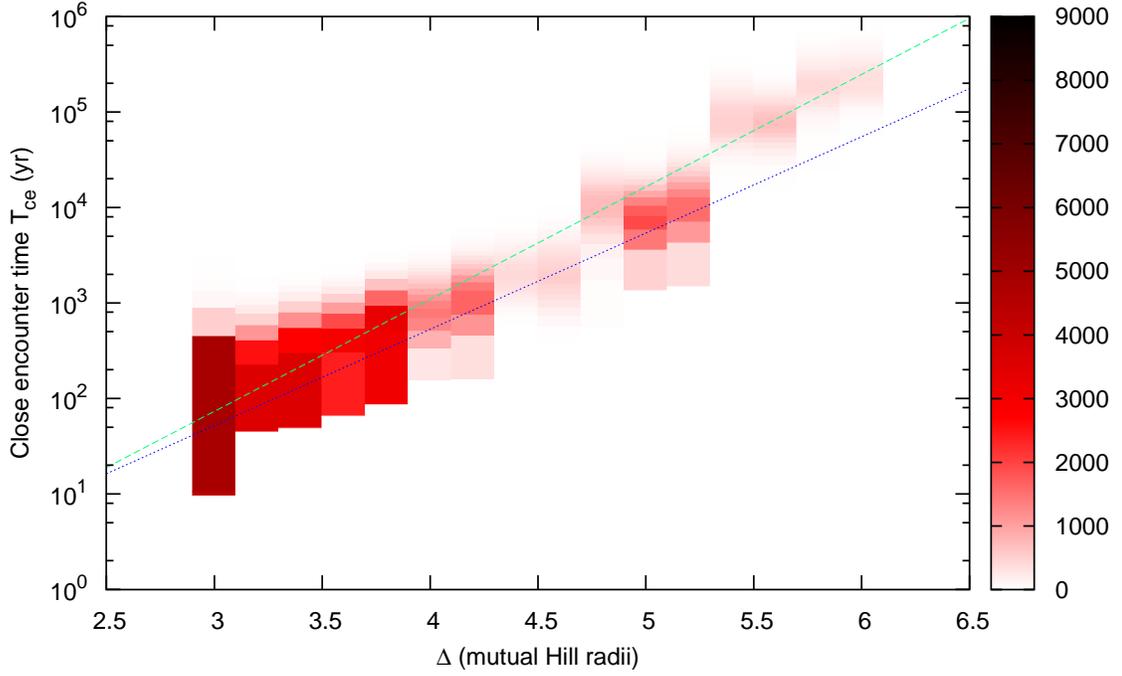}
    \caption{Histogram for models~1 through 16 in the plane of the initial
      orbital separation $\Delta$ and the duration of the
      pre-encounter phase, $T_{ce}$. The number of cases is expressed
      in color.  The blue dotted line is the fit to the peak 
(Eq.~(\ref{eq:delta_t})) and the green dashed line shows the $\Delta-T_{ce}$ 
relation in \citet{cwb96}.
}
    \label{delta_t}
    \end{center}
\end{figure}

In Fig.~\ref{delta_t}, we present a histogram in the $\Delta - T_{ce}$
plane expressed in color for the number of cases in our model calculations.
Connecting the peak, $T_{ce,peak}$, for each $\Delta$, we find the
following relation:
\begin{equation}\label{eq:delta_t}
\log T_{ce,peak} \simeq 1.008\Delta -1.307,
\end{equation}
which is very similar to what \citet{cwb96} found ($a=1.176,b=-1.663$
in their results) with a much smaller number ($5$) of realizations. 
It is clear from our results that the relation holds only for the
durations corresponding to the peaks and, in fact, the durations for each
$\Delta$ have a distribution as demonstrated above. Incidentally, the
integrals of motion, the linear and angular momenta and total energy,
are fixed in producing different realizations in this paper, which was
not the case for \citet{cwb96}.
 
%%%%%%%%%%%%%%%%%%%%%%%%%%%%%%%%%%%%%%%%%%%%%%%%%%%%%%%%%%%%%%%%%%%%%%%%%%%%%%%%%%%%%%%% 
\subsubsection{Jupiter system}
%%%%%%%%%%%%%%%%%%%%%%%%%%%%%%%%%%%%%%%%%%%%%%%%%%%%%%%%%%%%%%%%%%%%%%%%%%%%%%%%%%%%%%%%
One of the characteristics for the dynamics of the Jupiter systems is
that one of the planets are ejected from the system eventually. In
addition to the durations of the pre-encounter phase, $T_{ce}$, we
take also the statistics of the time from the first close encounter until
the ejection of a planet, $T_{ej} - T_{ce}$, for the Jupiter systems. 

\begin{figure}[h]
\begin{center}
    \plottwo{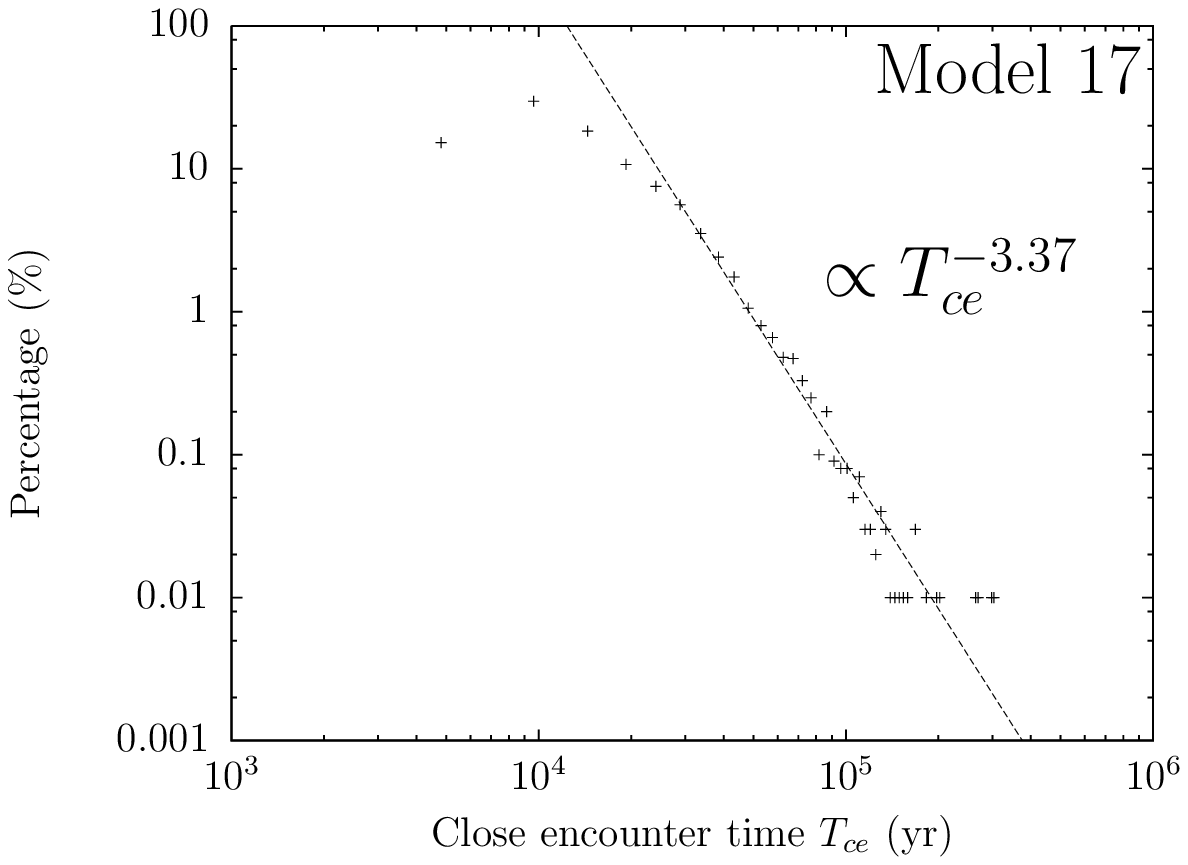}{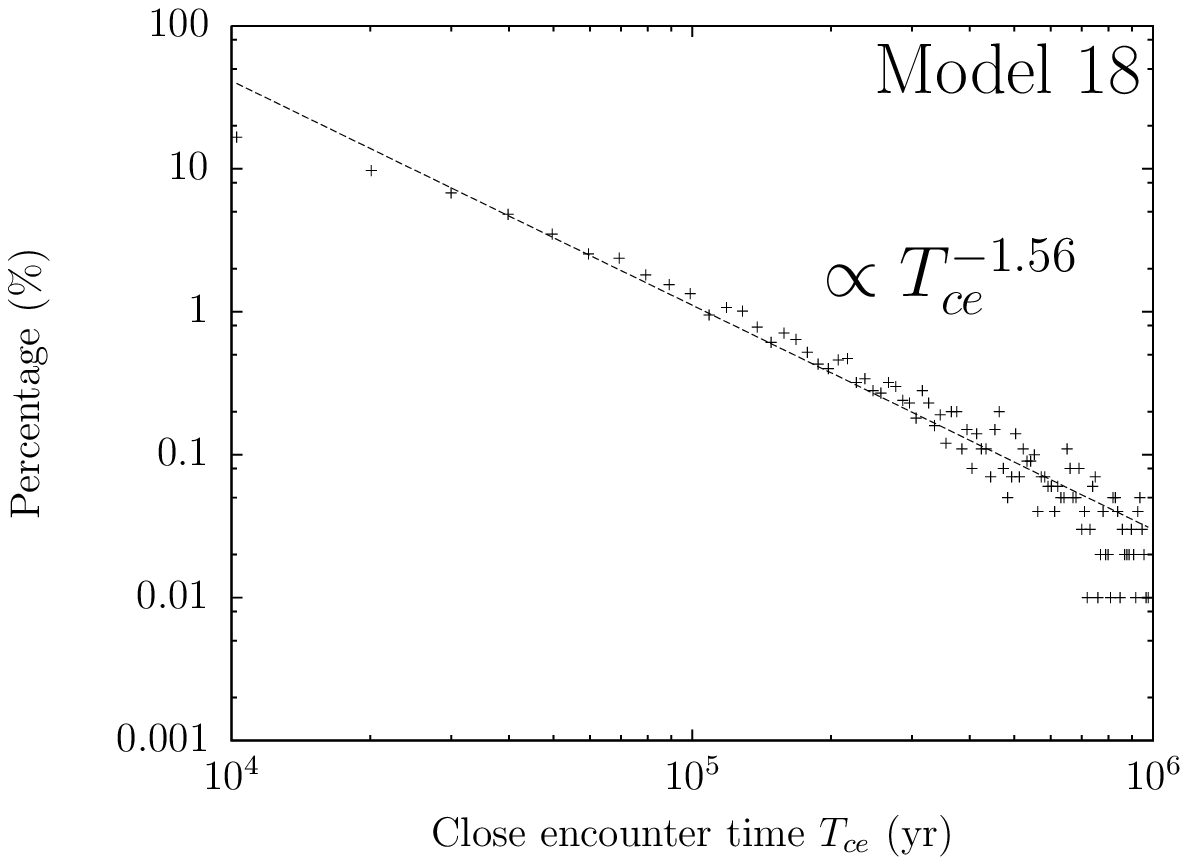}
    \caption{Normalized distributions of the duration of the
      pre-encounter phase, $T_{ce}$, for models~17 (left panel) and 18
      (right panel). 
      The straight lines are the fit by power law to the data in the 
      long-duration regime.
}
    \label{distbig1}
    \end{center}
\end{figure}

Fig.~\ref{distbig1} shows the distributions of $T_{ce}$ for models~17
(left panel) and 18 (right panel) for the Jupiter system.
Note that model~18 is meant to mimic the models in
\citet{mw02} and the initial condition is prepared differently from
other models. The straight lines in the figure are the power-law  fit
to the long-duration part of the distributions and obtained just in
the same manner as for the proto-planet systems (see
Fig.~\ref{distsmall}). It is seen again that the power-law
distribution is a good approximation in this regime, which suggests
that the quasi-regular motions before the close encounter in the 
Jupiter system can be also understood as the stagnant motion just as 
for the proto-planet system. 
The obtained power-law indices are 
$-5.57$ and $-1.56$ for models~17 and 18, respectively.

\begin{figure}[h]
\begin{center}
\plottwo{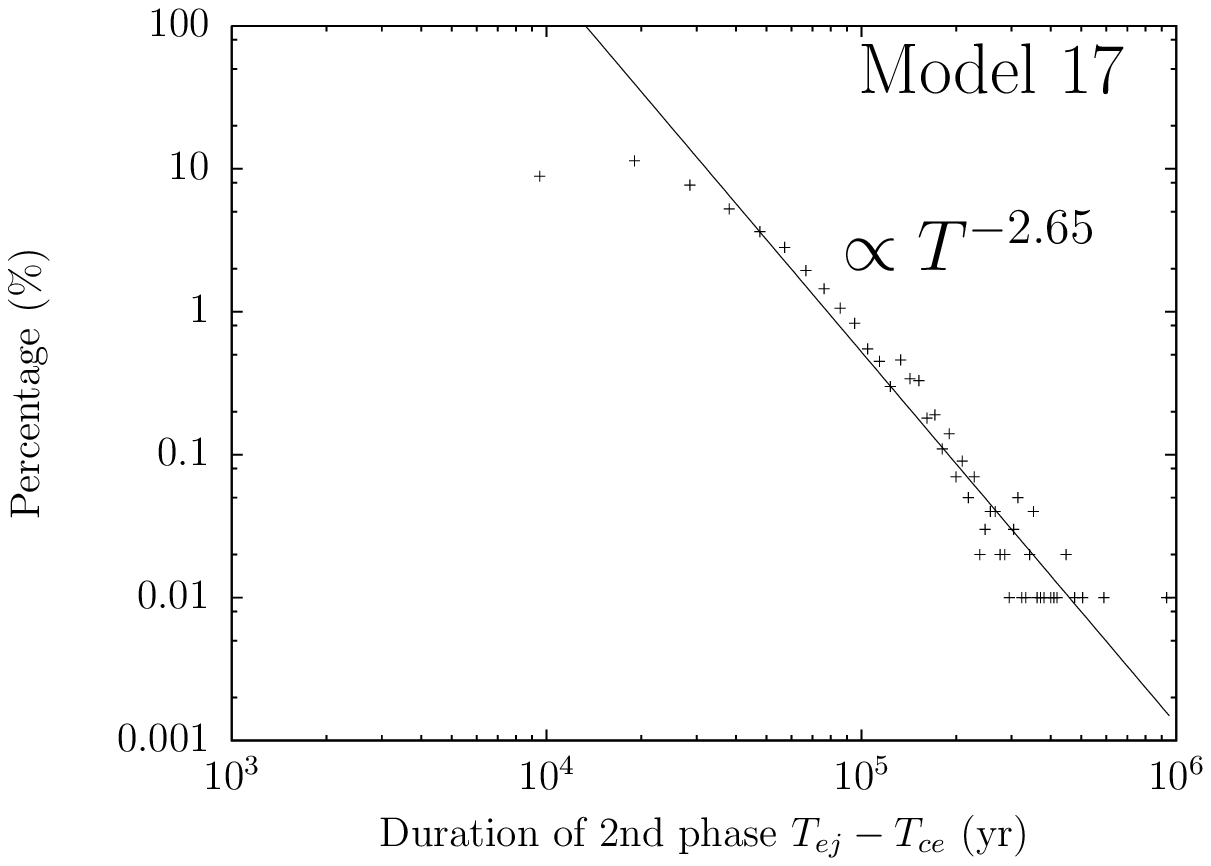}{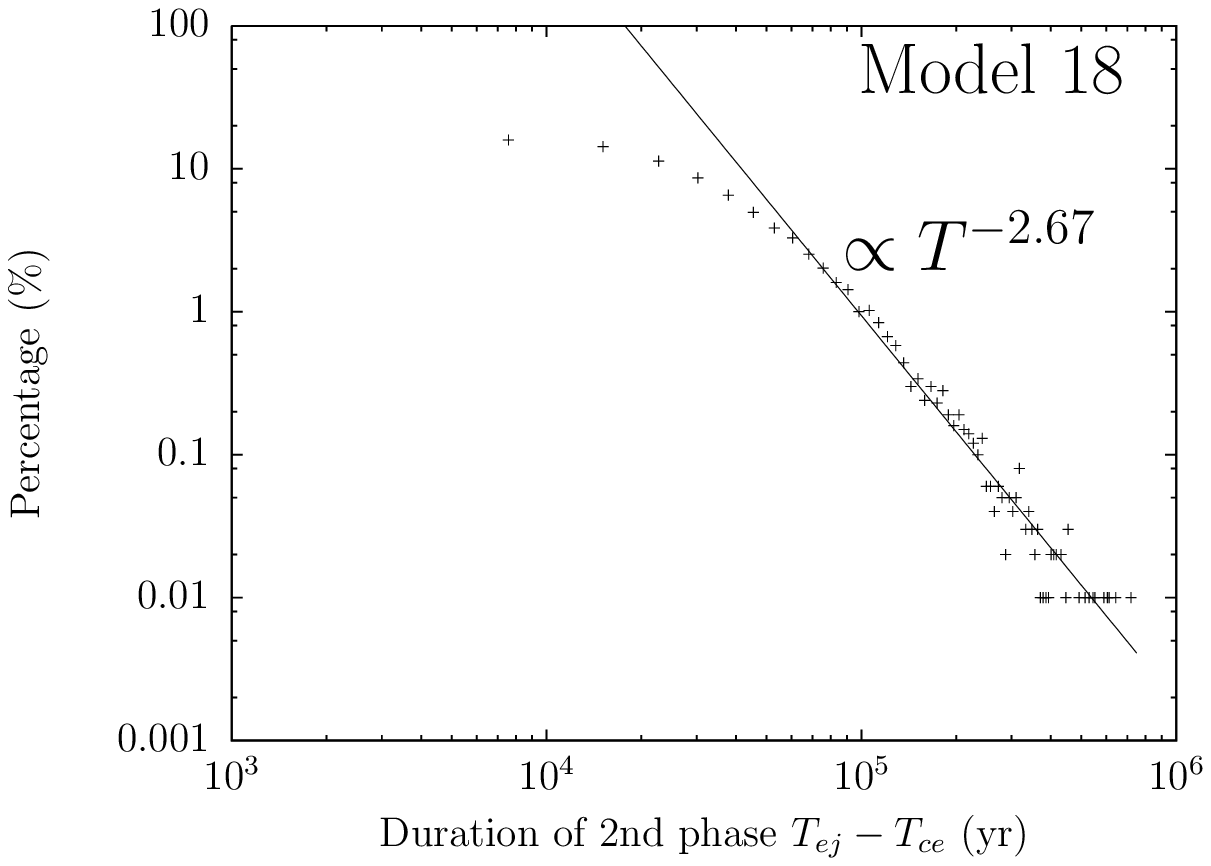}
    \caption{Normalized distributions of the time from the first close 
encounter up to the ejection of a planet, $T_{ej}-T_{ce}$,
for models~17 (left panel) and 18 (right panel). The straight lines
are the fit by power law to the data in the long-time regime.
}
    \label{distbig2}
    \label{lastfig}
    \end{center}
\end{figure}

Now we turn attention to the distributions of the time from the
encounter to the ejection, $T_{ej}-T_{ce}$, which are given in 
Fig.~\ref{distbig2} for models~17 (left panel) and 18 (right panel).
It is clear from the figure that the long-time portions of both
distributions are again approximated by the power laws with the 
indices of $-2.65$ and $-2.67$ for models~17 and 18, respectively.
This implies that there is no characteristic time scale for the
duration, during which the phase space orbit of the system remains 
in the part of the phase space corresponding to the bound states of
three planets.

If the power-law distribution of $T_{ce}$ reflects 
the self-similar distribution of smaller tori in the stagnant layer of
a KAM torus (see Fig.~\ref{schematic1}) as claimed in the stagnant 
motion model, the power-law distribution of $T_{ej}-T_{ce}$ might
suggest a self-similar distribution of these KAM tori in the 
part of the phase space for the bound motions of three planets 
(see Fig.~\ref{schematic2} for a schematic picture of the phase space).
% If the power-law distribution of $T_{ce}$ reflects 
% the self-similar distribution of smaller tori in the stagnant layer of
% a KAM torus (see Fig.~\ref{schematic1}) as claimed in the stagnant 
% motion model, the power-law distribution of $T_{ej}-T_{ce}$ might 
% suggest a self-similar distribution of these KAM tori in the 
% part of the phase space for the bound motions of three planets 
% (see Fig.~\ref{schematic2} for a schematic picture of the phase space).
%%%%%%%%%%%%%%%%%%%%%%%%%%%%%%%%%%%%%%%%%%%%%%%%%%%%%%%%%%%%%%%%%%%%%%%%%%%%%%%%%%%%%%%%
\section{Summary and Discussions}
%%%%%%%%%%%%%%%%%%%%%%%%%%%%%%%%%%%%%%%%%%%%%%%%%%%%%%%%%%%%%%%%%%%%%%%%%%%%%%%%%%%%%%%%

In this paper we have numerically investigated the dynamics of the
three-planet system and inferred its phase space structure from the
obtained Lyapunov exponents, power spectra of orbital elements and
distributions of induction periods based on the stagnant motion model.
What we have found are the followings:

1. The global and local Lyapunov exponents show that the system is
almost non-chaotic until the first close encounter between two planets 
and it then turns into chaotic motions with intermittent non-chaotic
periods. This suggests that the phase space consists of KAM tori surrounded by 
the stagnant layer and immersed in the chaotic sea. In fact, the dense
bands are formed in the action-angle plot, corresponding to the quasi-regular 
motions. The lines composing a band are undulating with frequencies that
obey a power-law and this may represent the motions around the satellite tori in the stagnant layer,
which have a self-similar distribution. The phase space orbit  goes
from one system of KAM torus and stagnant layer to another through the chaotic sea.
% Incidentally, it is interesting to 
% mention that the time variation of local Lyapunov exponent shows that 
% the system is almost non-chaotic just prior to the ejection of a
% planet in the Jupiter system.

2. The power spectra of the orbital eccentricities of planets can be
approximated by the power law, $\propto 1/f^{\nu}$, in general. 
Such power-law spectra are known to be one of the characteristic features of 
the stagnant motions although the power-law index is not predicted by the theory. In our models
the power-law index is $\nu \simeq 1$ for the pre-encounter phase whereas it
becomes $\nu \simeq 2$ after the encounter.
The spectrum in the post-encounter phase 
is similar to that of the Brownian motions or the random walks.
On the other hand, the spectrum in the pre-encounter phase might be originated
from the fractional Brownian motions.
%  It should be noted that
% the former is one of the features that characterize the stagnant
% motion. On the other hand, the spectrum in the post-encounter phase 
% is similar to that of the Brownian motions or the random walks. Hence the 
% deviation from the initial regular motions grows more slowly than the
% random walk until the first close encounter. 

3. The distributions of the duration of the pre-encounter phase that
was referred to as the induction periods obey the power law in the
long-duration part. The power-law indices are substantially different
between models. It is stressed that the stagnant motion model predicts
the power law for the distribution of the induction periods as a
consequence of the self-similar distribution of smaller tori in the
stagnant layer around a KAM torus. The distributions are deviated
from the power law in the short-duration part and has a peak in
between. Connecting the peaks for various models with different initial
orbital separations, we have obtained the relation similar to what  
\citet{cwb96} found. It is also shown that the duration of the pre-encounter phase
has actually a considerably broad distribution.

\begin{figure}
\begin{center}
 \resizebox{140mm}{!}{\plotone{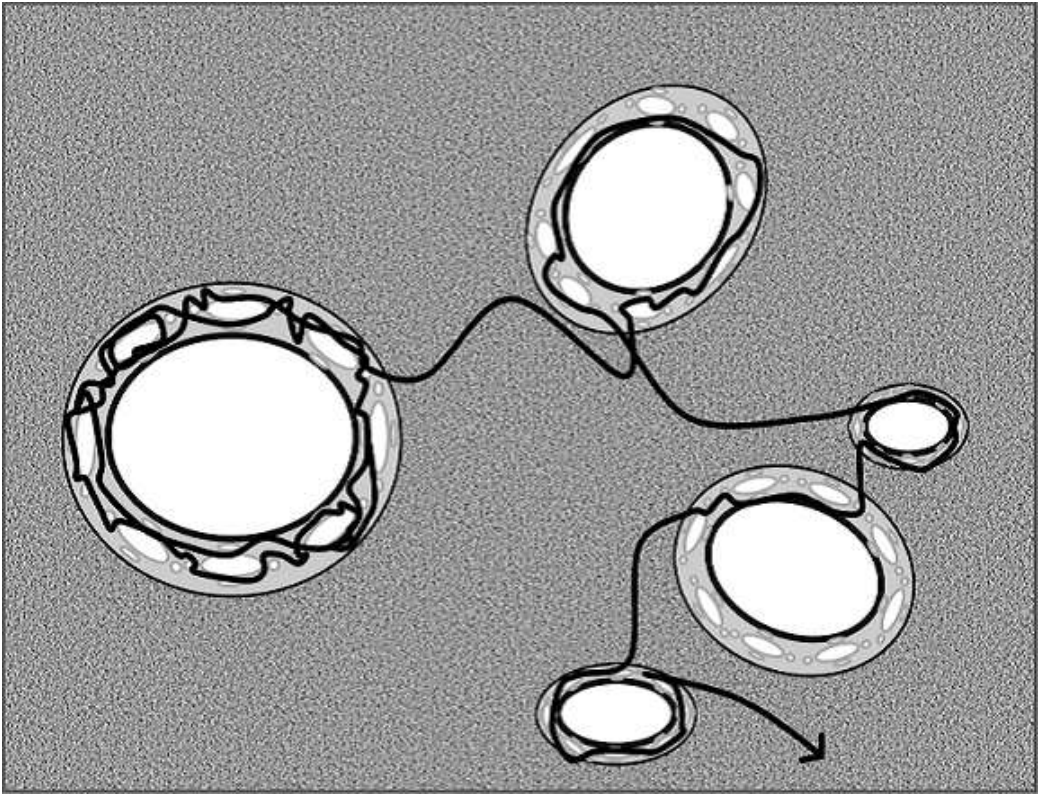} }
\caption{A schematic illustration of the phase space suggested by the
  power-law distributions of the time from the encounter to the
  ejection of a planet for the Jupiter system.
%  KAM tori with stagnant layers and chaotic region are distributed in laminae.
 KAM tori of various sizes with a stagnant layer and smaller tori in
 it are distributed self-similarly in the chaotic sea. An exemplary
 phase space orbit is given by a curve with an arrow.
}
% \caption{A schematic illustration of the phase space suggested by the
%   power-law distributions of the time from the encounter to the
%   ejection of a planet for the Jupiter system.
%  KAM tori of various sizes with a stagnant layer and smaller tori in
%  it are distributed self-similarly in the chaotic sea. An exemplary
%  phase space orbit is given by a curve with an arrow.
% }
    \label{schematic2}		
    \end{center}
\end{figure}

4. For the Jupiter system, the distribution of the time from the first
encounter to the ejection of a planet from the system also obeys a
power law, which was not expected initially. 
From the analogy to the stagnant
motion model, we might be able to infer the phase space structure as 
shown schematically in Fig.~\ref{schematic2}: Many KAM tori with its
own stagnant layer and satellite tori in it are distributed
self-similarly in the chaotic sea.

Although we expect the phase space structure depicted in
Fig.~\ref{schematic2} is true both of the proto-planet system and 
the Jupiter system, the difference between them should be also mentioned.
In general, the number of KAM tori in the phase space becomes smaller 
and the stagnant layers around them get thinner as the perturbation
to the integrable system is greater. In the system of our concern, 
the perturbation is the interactions among the planets and hence 
it is larger for more massive planets. The pre-encounter phase is an
exception, though. In this phase, the planets have nearly circular 
orbits separated by several Hill radii. Then the strength of the 
interactions between the planets depends only weakly on the planetary 
mass thanks to the definition of the Hill radius given in Eq.~(\ref{eq:hill}).
These facts suggest that the KAM torus corresponding to the initial
regular motion and its stagnant layer are robust and similar for the
proto-planet and Jupiter systems while the number of other KAM tori
is smaller and their stagnant layers are thinner for the Jupiter
system than for the proto-planet system. This difference in the phase 
space structures is supposed to be responsible for the observed
difference in the orbital evolutions of the two systems: one of 
the planets is ejected in short times for the Jupiter system whereas
no ejection occurs at least for $\sim 10^7$ years in the proto-planet 
system.

The results obtained in this paper appear to be consistent with our
interpretation that the dynamics of three-planet system is a stagnant
motion at least in the pre-encounter phase. It is also suggested from 
the results for the Jupiter system that even the post-encounter phase may be 
described by some extension of the stagnant motion model.
It is true, however, that a more direct capture of satellite tori in the phase space is certainly desirable. 
Although we have attempted to do this with the so-called Poincare mapping, but in vain 
so far. We are afraid that the degree of freedom of our system is just too large to find
an appropriate two dimensional section in the 12-dimensional phase space. Maybe other approaches
should be pursued in the future work. In so doing, the number of realizations should be increased 
and other initial settings should be tried. Not to mention, we are also interested in how 
the results will change as the number of planets are varied.

%%%%%%%%%%%%%%%%%%%%%%%%%%%%%%%%%%%%%%%%%%%%%%%%%%%%%%%%%%%%%%%%%%%%%%%
% \newpage
\acknowledgements

\begin{center}
  {\bf Acknowledgments}
\end{center}
We would like to thank Shigeru Ida and Tetsuro Konishi for their useful comments. 
This work was partially supported by Grants-in-Aid for the Scientific
Research from the Ministry of Education, Science and
Culture of Japan (17540267, 19104006) and the 21st-Century COE Program
``Holistic Research and Education Center for Physics of 
Self-organization Systems.''

%%%%%%%%%%%%%%%%%%%%%%%%%%%%%%%%%%%%%%%%%%%%%%%%%%%%%%%%%%%%%%%%%%%%%%%

%%%%%%%%%%%%%%%%%%%%%%%%%%%%%%%%%%%%%%%%%%%%%%%%%%%%%%%%%%%%%%%%%%%%%%%%
%%%%%%%%%%%%%%%%%%%%%%%%%%%%%%%%%%%%%%%%%%%%%%%%%%%%%%%%%%%%%%%%%%%%%%%%

\clearpage

%%%%%%%%%%%%%%%%%%%%%%%%%%%%%%%%%%%%%%%%%%%%%%%%%%%%%%%%%%%%%%%%%%%%%%%
%%%%%%%%%%%%%%%%%%%%%%%%%%%%%%%%%%%%%%%%%%%%%%%%%%%%%%%%%%%%%%%%%%%%%%%

%%%%%%%%%%%%%%%%%%%%%%%%%%%%%%%%%%%%%%%%%%%%%%%%%%%%%%%%%%%%%%%%%%%%%%%

%%%%%%%%%%%%%%%%%%%%%%%%%%%%%%%%%%%%%%%%%%%%%%%%%%%%%%%%%%%%%%%%%%%%%%

%%%%%%%%%%%%%%%%%%%%%%%%%%%%%%%%%%%%%%%%%%%%%%%%%%%%%%%%%%%%%%%%%%%%%%%

%%%%%%%%%%%%%%%%%%%%%%%%%%%%%%%%%%%%%%%%%%%%%%%%%%%%%%%%%%%%%%%%%%%%%%%
%%%%%%%%%%%%%%%%%%%%%%%%%%%%%%%%%%%%%%%%%%%%%%%%%%%%%%%%%%%%%%%%%%%%%%%
%%%%%%%%%%%%%%%%%%%%%%%%%%%%%%%%%%%%%%%%%%%%%%%%%%%%%%%%%%%%%%%%%%%%%%%
%%%%%%%%%%%%%%%%%%%%%%%%%%%%%%%%%%%%%%%%%%%%%%%%%%%%%%%%%%%%%%%%%%%%%%%

%%%%%%%%%%%%%%%%%%%%%%%%%%%%%%%%%%%%%%%%%%%%%%%%%%%%%%%%%%%%%%%%%%%%%%%

%%%%%%%%%%%%%%%%%%%%%%%%%%%%%%%%%%%%%%%%%%%%%%%%%%%%%%%%%%%%%%%%%%%%%%%

\begin{figure}
\begin{center}
    \begin{tabular}{ll}
      \plottwo{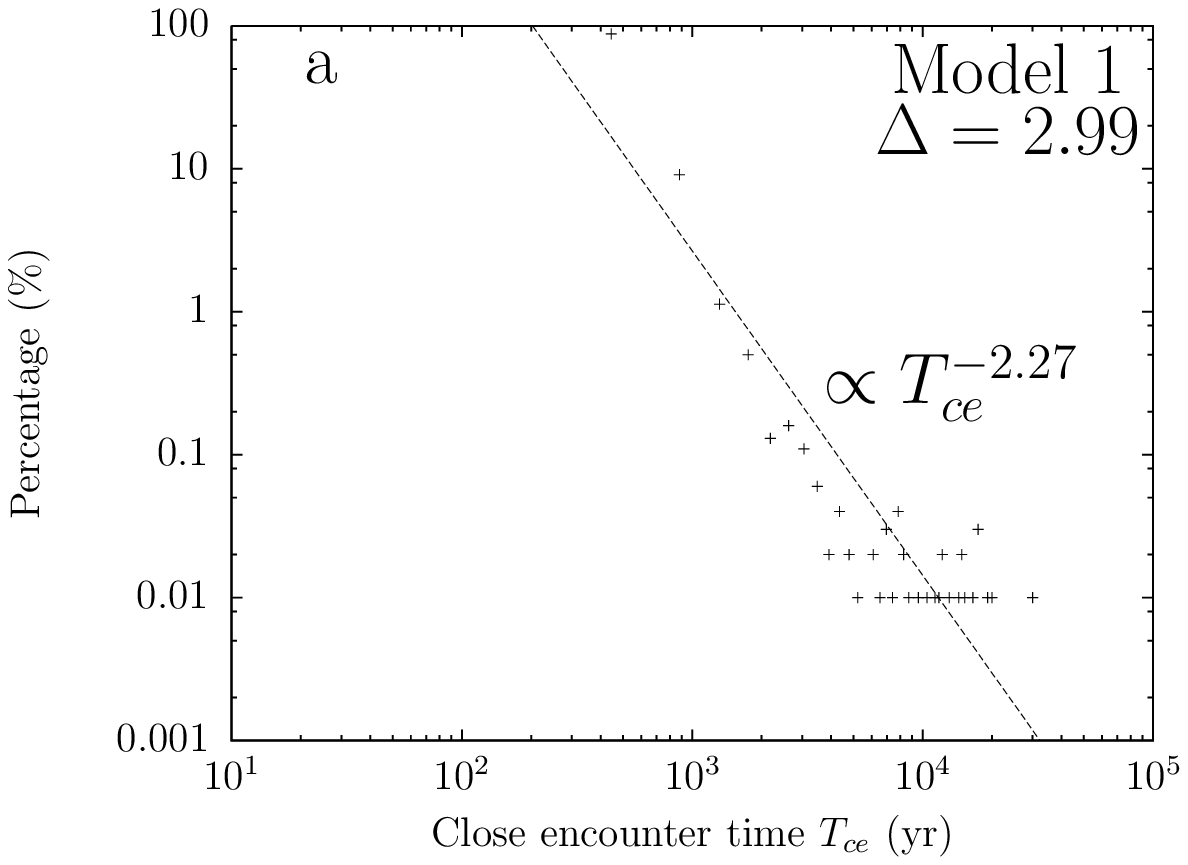}{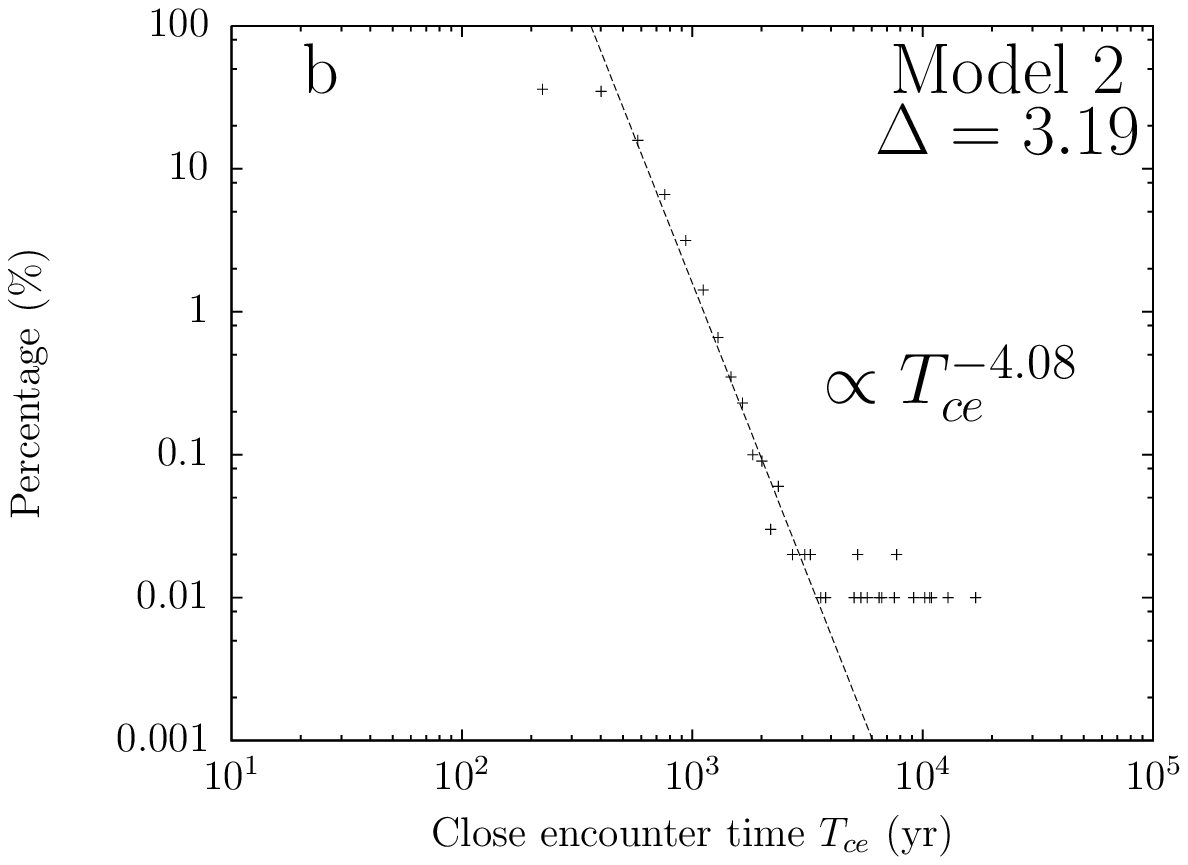} \\
      \plottwo{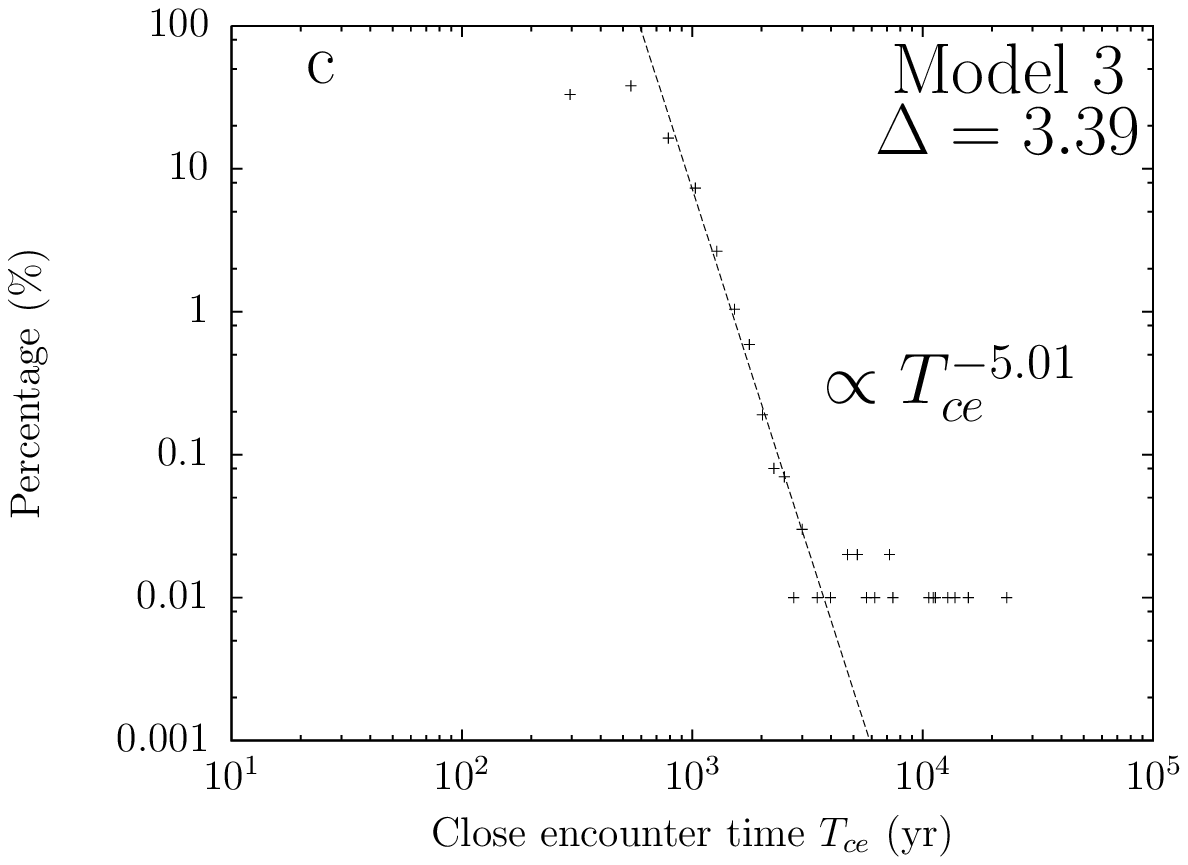}{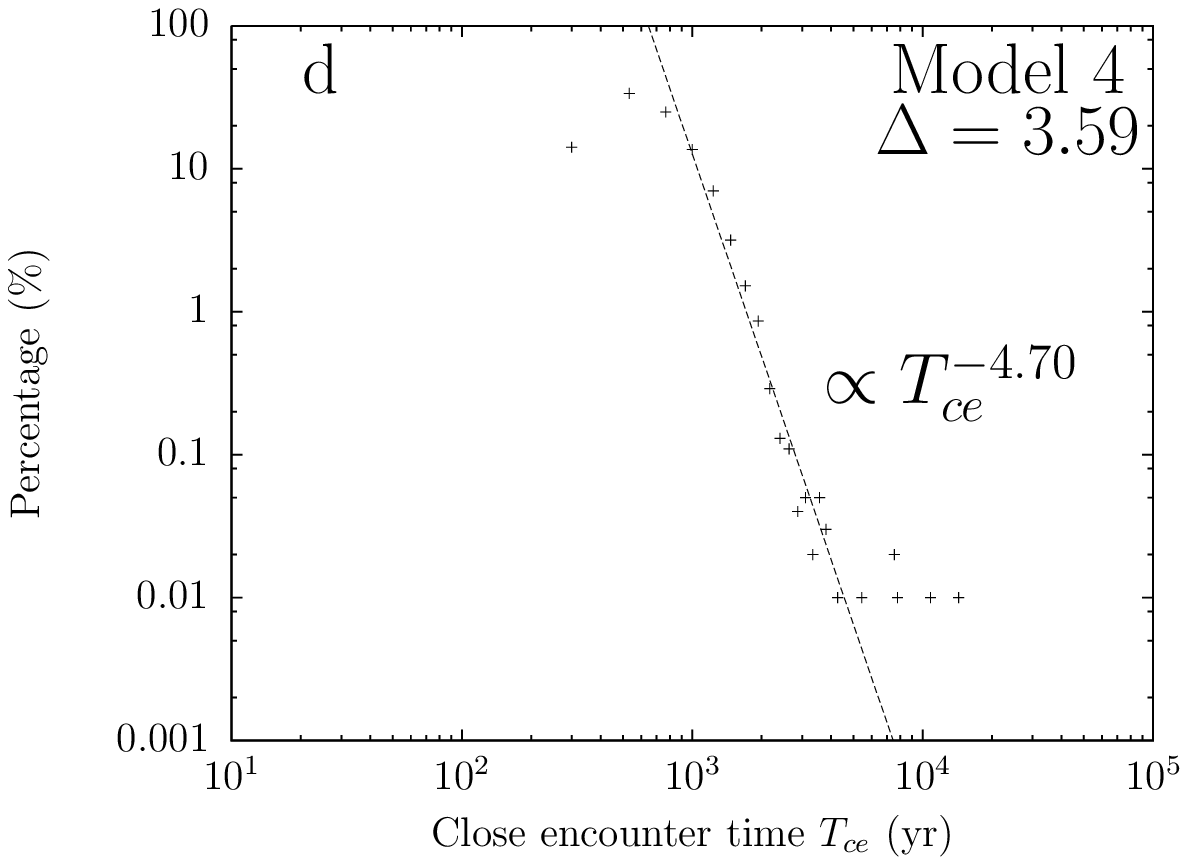} \\
      \plottwo{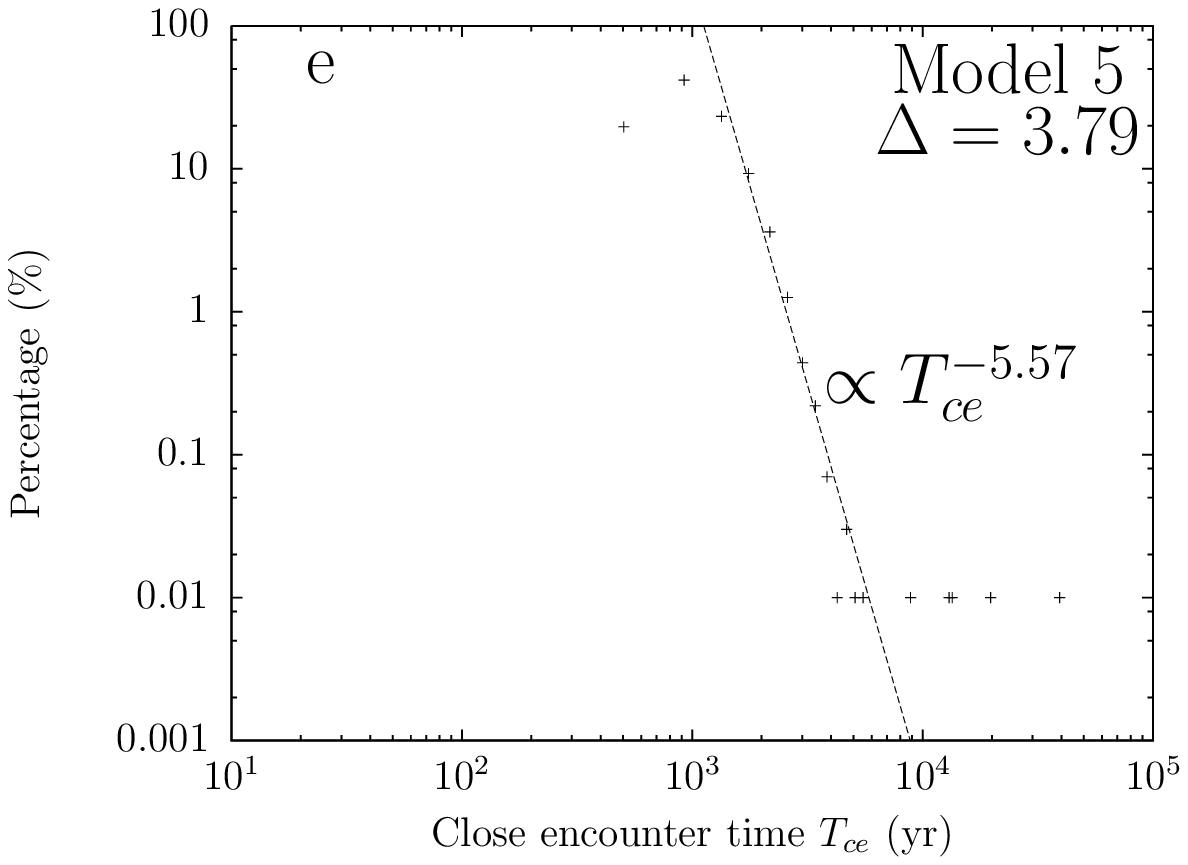}{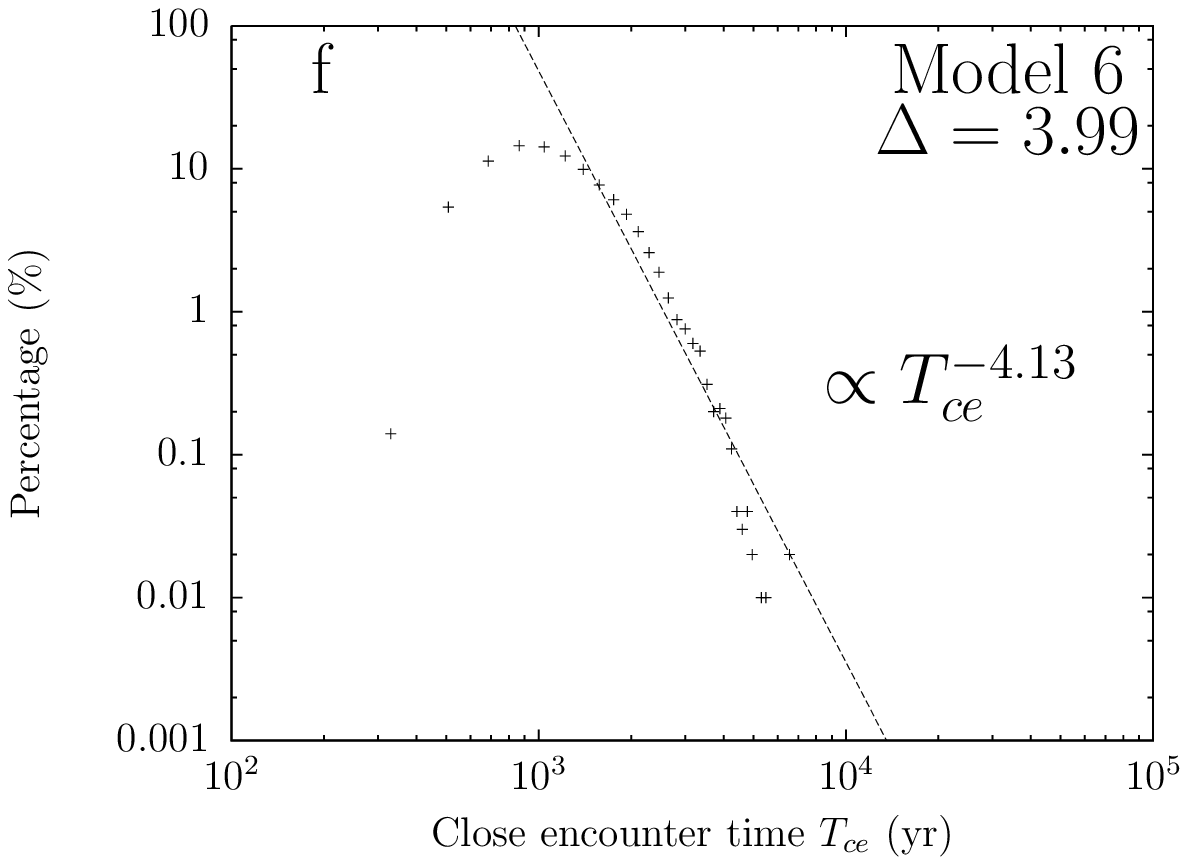} \\
    \end{tabular}
   \end{center}
\end{figure}

%%%%%%%%%%%%%%%%%%%%%%%%%%%%%%%%%%%%%%%%%%%%%%%%%%%%%%%%%%%%%%%%%%%%%%%
\begin{figure}
\begin{center}
\begin{tabular}{ll}
      \plottwo{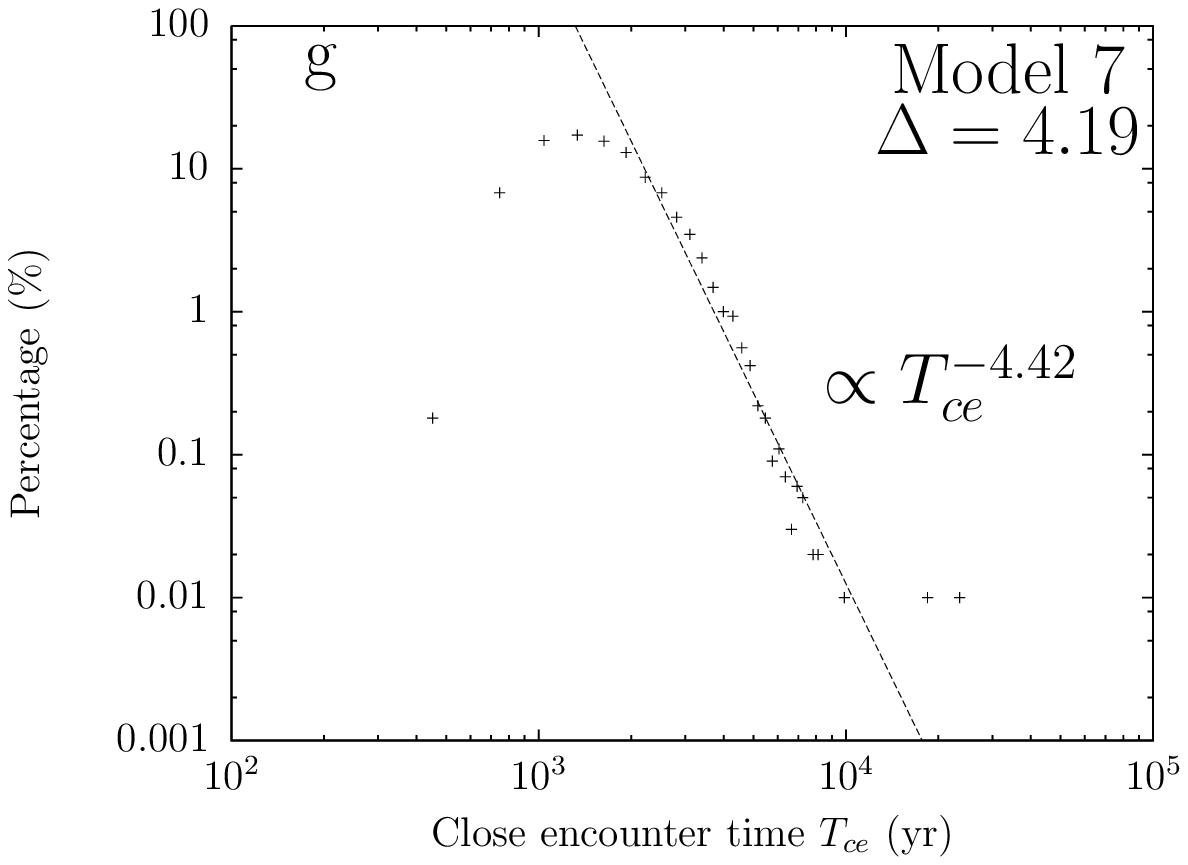}{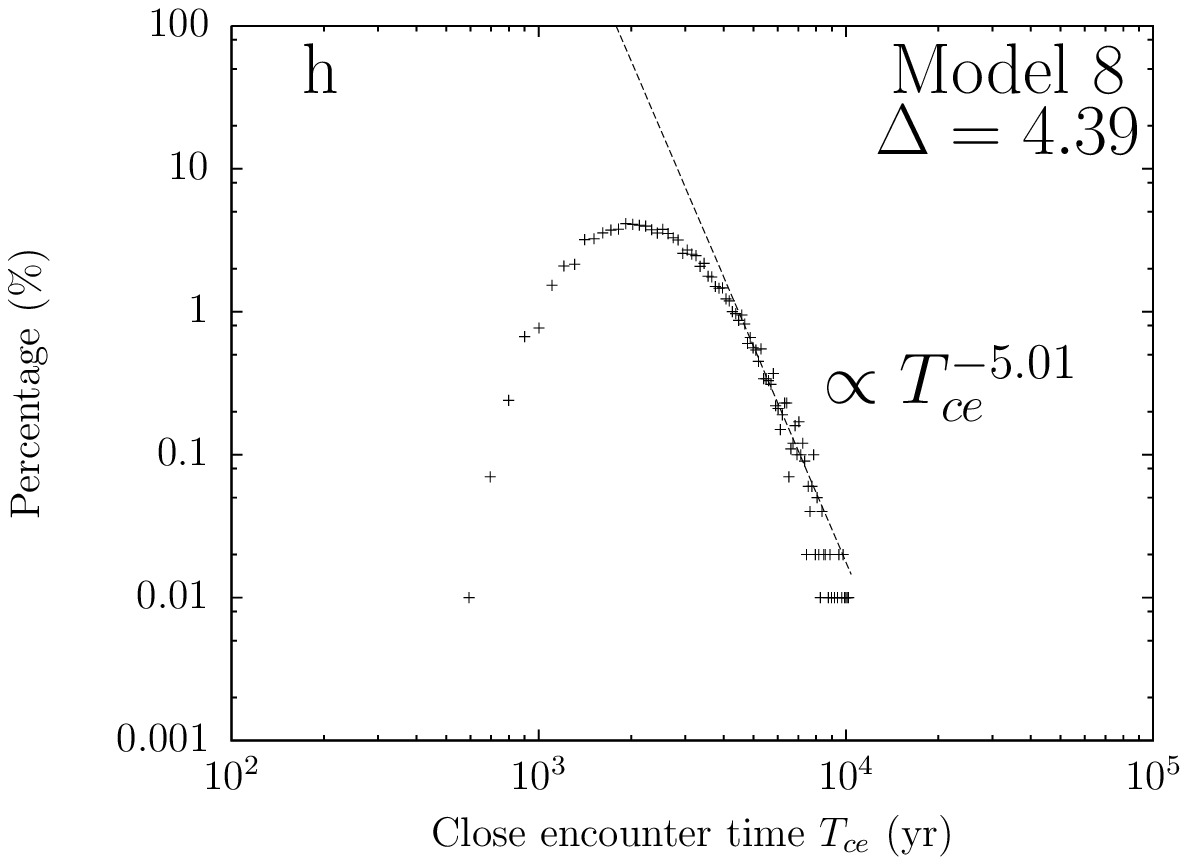} \\
      \plottwo{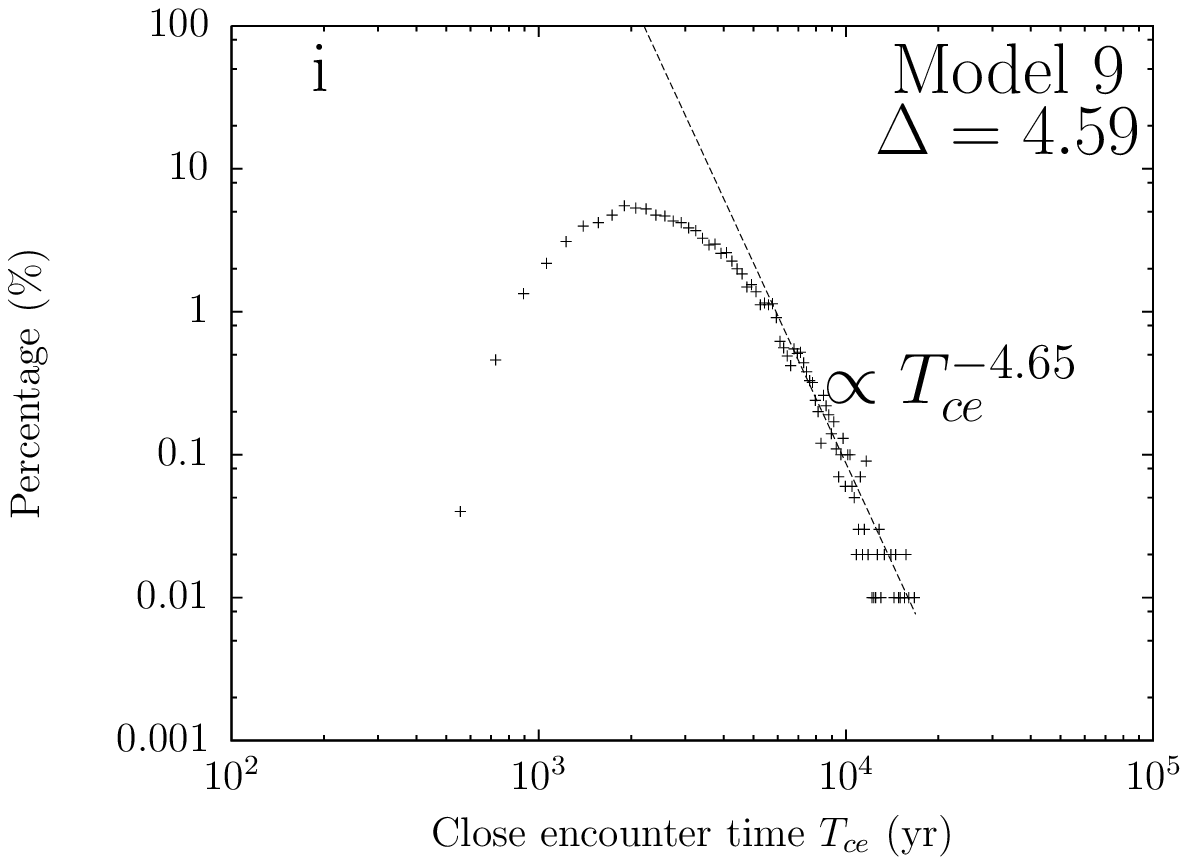}{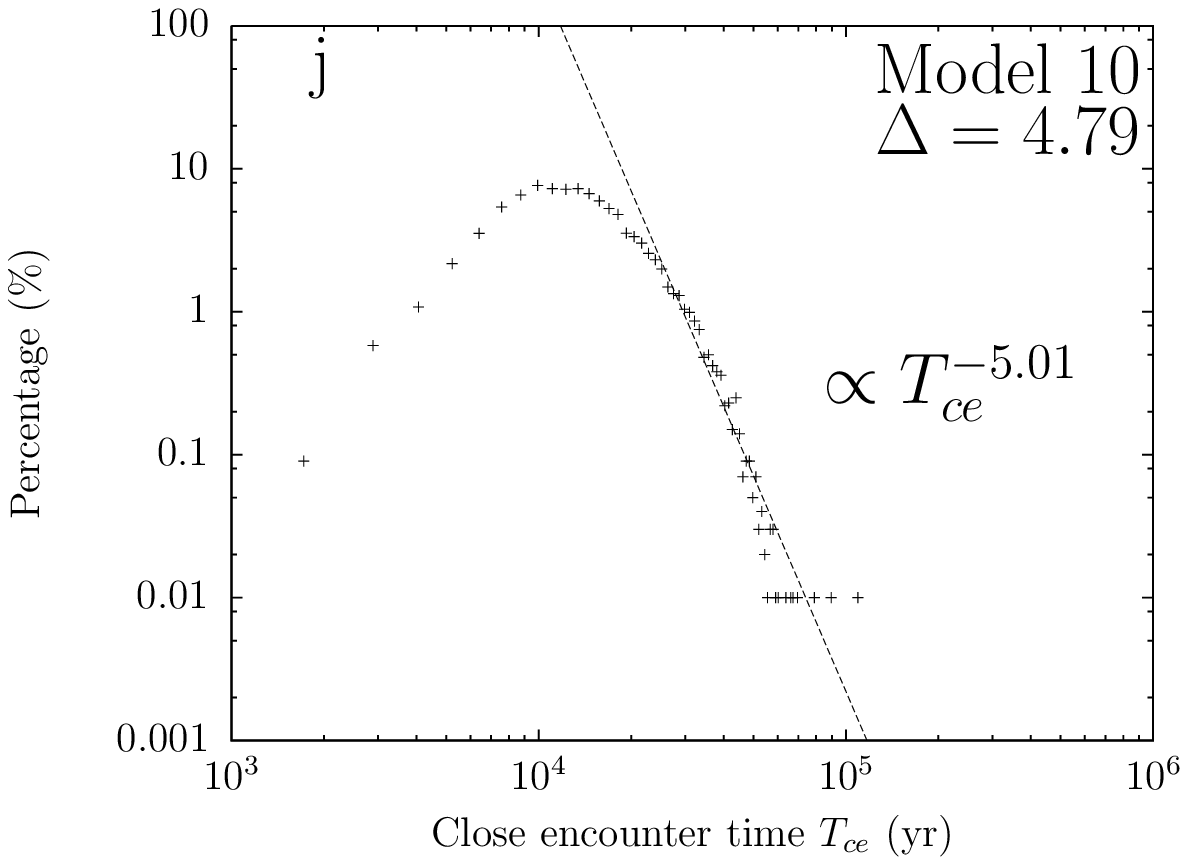} \\
      \plottwo{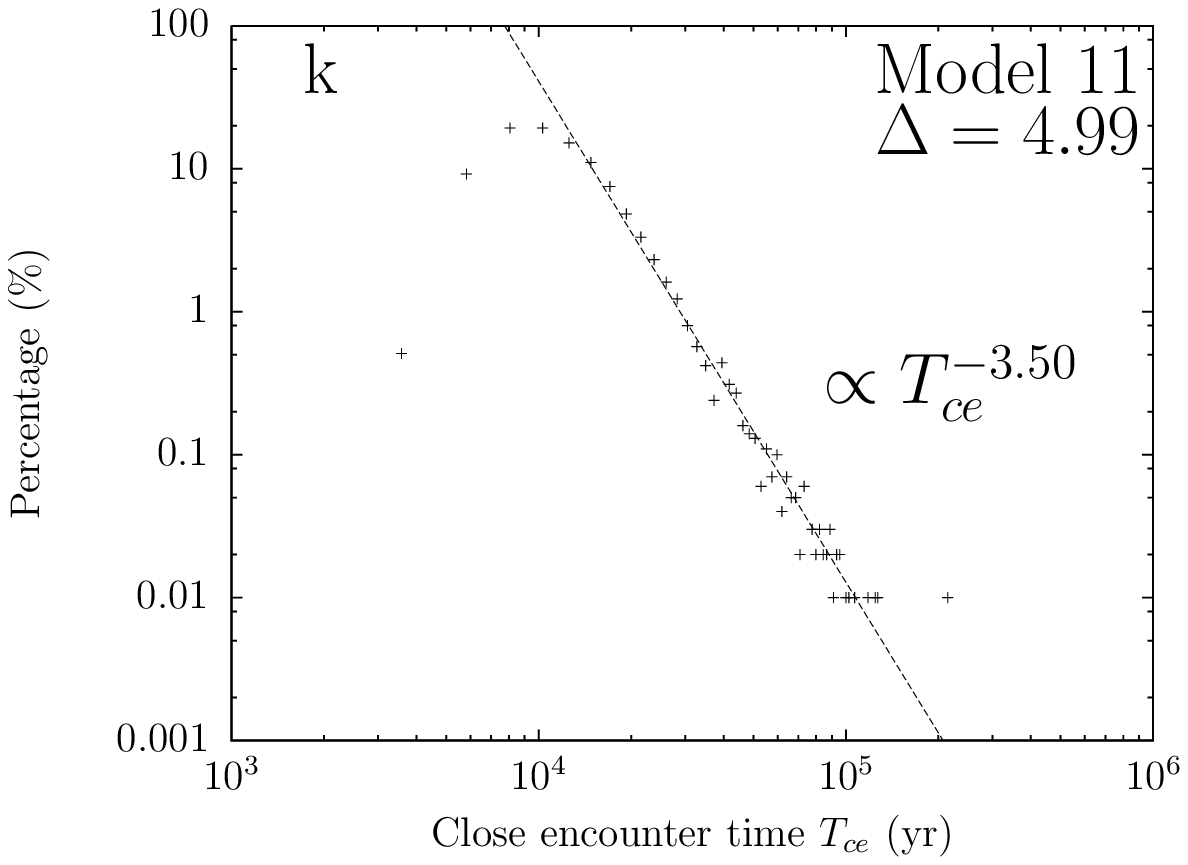}{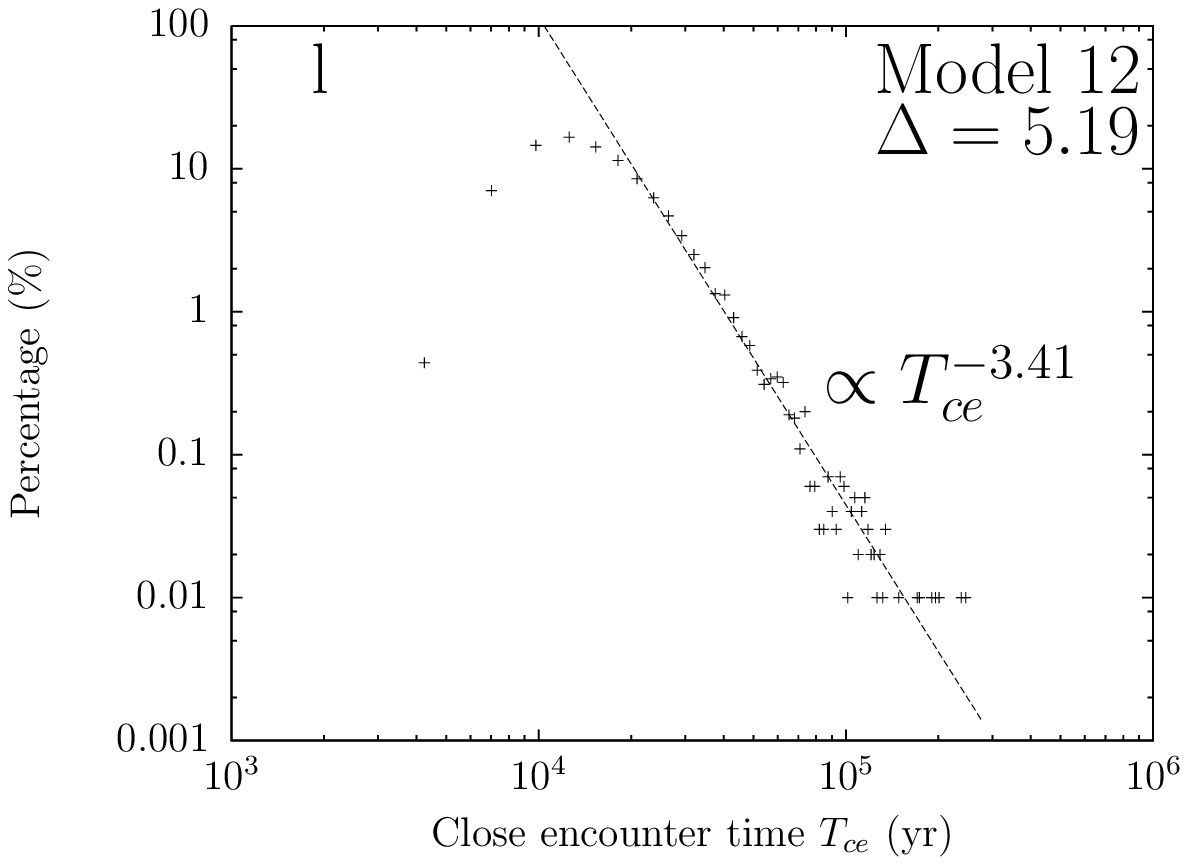} \\
    \end{tabular}
\end{center}
\end{figure}
%%%%%%%%%%%%%%%%%%%%%%%%%%%%%%%%%%%%%%%%%%%%%%%%%%%%%%%%%%%%%%%%%%%%%%%
%%%%%%%%%%%%%%%%%%%%%%%%%%%%%%%%%%%%%%%%%%%%%%%%%%%%%%%%%%%%%%%%%%%%%%%

%%%%%%%%%%%%%%%%%%%%%%%%%%%%%%%%%%%%%%%%%%%%%%%%%%%%%%%%%%%%%%%%%%%%%%%
%%%%%%%%%%%%%%%%%%%%%%%%%%%%%%%%%%%%%%%%%%%%%%%%%%%%%%%%%%%%%%%%%%%%%%%

%%%%%%%%%%%%%%%%%%%%%%%%%%%%%%%%%%%%%%%%%%%%%%%%%%%%%%%%%%%%%%%%%%%%%%%

%%%%%%%%%%%%%%%%%%%%%%%%%%%%%%%%%%%%%%%%%%%%%%%%%%%%%%%%%%%%%%%%%%%%%%%
% \begin{deluxetable}{lc}
% \tabletypesize{\scriptsize}
% \tabletypesize{\footnotesize}
% \tablecaption{\textcolor{red}{Models}}
% \tablewidth{0pt}
% \tablehead{
% \colhead{Planetary mass $m_{pl}$ ($M_\odot$)} & 
%  \colhead{Radial location of planets} }
%  
% \startdata
% $10^{-7}$  & $a_1=1$\hspace{3mm}(AU),\hspace{2mm} $\Delta=3.0 - 6.0$ \\
% $10^{-3}$  & $a_1=5$\hspace{3mm}(AU),\hspace{2mm} $\Delta=3.4$ \\
% $10^{-3}$  & $a_1=5$, \hspace{3mm}$a_2=7.25$,\hspace{3mm}$a_3=9.5$\hspace{3mm}(AU)\\
% \enddata

% \tablecomments{ The models we investigated. In the models denoted in upper two lines we set planets following to
%  Eq.(\ref{eq:initial}). $\Delta$ denotes the orbital separation of planets measured in Hill radius.
% For simplicity, in the all the system we restrict planets to have equal mass and coplanar, circular orbits.
% We make $10000$ models for each model in this table to compute the distribution of induction periods.
% }
% \label{models}
% \end{deluxetable}
%%%%%%%%%%%%%%%%%%%%%%%%%%%%%%%%%%%%%%%%%%%%%%%%%%%%%%%%%%%%%%%%%%%%%%%
% \clearpage
% \thispagestyle{empty}
\begin{deluxetable}{ccccccccc} 
\tabletypesize{\tiny}
\tablecaption{Model parameters}
\tablewidth{0pt}
\tablehead{
  \colhead{Model}& \colhead{$m_{pl}$} & \colhead{$a_1$} & \colhead{$E_{tot}$} & \colhead{$L_{ztot}$ } & \colhead{$P_{xtot}$}
 & \colhead{$P_{ytot}$} & \colhead{ $<\Delta> \hspace{3mm} (\sigma^2 _{\Delta})$} & $<e> \hspace{3mm} (\sigma^2 _{e})$ \\
    \colhead{}& \colhead{($M_\odot$)} & (AU) & \colhead{ ( $\times 10^{39}$erg)} & \colhead{($ \times 10^{46}\mathrm{ g cm^2 s^{-1}}$)} & \colhead{($\times 10^{30}\mathrm{ cm s^{-1} }$)}
 & \colhead{($\times 10^{30} \mathrm{cm s^{-1}}$)} & \colhead{(Hill radius)} &  \\
 & & & & & & & \colhead{ \tablenotemark{a} $<a_2>\hspace{1mm}(\sigma^2 _{a_2}),\hspace{1mm}<a_3>\hspace{1mm}(\sigma^2 _{a_3})$} &  \\
 & & & & & & & (AU) & \\
}
\rotate
\clearpage
\startdata
1 & $10^{-7}$ & 1 & -2.6152  & 2.6752  & -3.0924  & 5.3780 & $2.9970  \hspace{3mm}( 1.3965\times 10^{-4})$ & $1.2129\times 10^{-7} \hspace{3mm}(3.7549\times 10^{-15})$ \\
2 & $10^{-7}$ & 1 & -2.6131  & 2.6763  & -3.2966  & 5.7347 & $3.1973  \hspace{3mm}( 1.1031\times 10^{-4})$ & $1.2123\times 10^{-7} \hspace{3mm}(3.7505\times 10^{-15})$ \\
3 & $10^{-7}$ & 1 & -2.6110  & 2.6774  & -3.5005  & 6.0910 & $3.3975  \hspace{3mm}( 8.9031\times 10^{-5})$ & $1.2118\times 10^{-7} \hspace{3mm}(3.7451\times 10^{-15})$ \\
4 & $10^{-7}$ & 1 & -2.6089  & 2.6785  & -3.7042  & 6.4472 & $3.5977  \hspace{3mm}( 7.3037\times 10^{-5})$ & $1.2112\times 10^{-7} \hspace{3mm}(3.7404\times 10^{-15})$ \\
5 & $10^{-7}$ & 1 & -2.6068  & 2.6796  & -3.9076  & 6.8031 & $3.7979  \hspace{3mm}( 6.0711\times 10^{-5})$ & $1.2105\times 10^{-7} \hspace{3mm}(3.7370\times 10^{-15})$ \\
6 & $10^{-7}$ & 1 & -2.6047  & 2.6806  & -4.1108  & 7.1587   & $3.9980  \hspace{3mm}( 5.1030\times 10^{-5})$ & $1.2100\times 10^{-7} \hspace{3mm}(3.7317\times 10^{-15})$ \\
7 & $10^{-7}$ & 1 & -2.6026  & 2.6817  & -4.3137  & 7.5142  & $4.1982  \hspace{3mm}( 4.3308\times 10^{-5})$ & $1.2094\times 10^{-7} \hspace{3mm}(3.7274\times 10^{-15})$ \\
8 & $10^{-7}$ & 1 & -2.6005  & 2.6828  & -4.5164  & 7.8693  & $4.3983  \hspace{3mm}( 3.7065\times 10^{-5})$ & $1.2088 \times 10^{-7} \hspace{3mm}(3.7223\times 10^{-15})$ \\
9 & $10^{-7}$ & 1 & -2.5985  & 2.6839  & -4.7188  & 8.2243 & $4.5984  \hspace{3mm}( 3.1962\times 10^{-5})$ & $1.2082\times 10^{-7} \hspace{3mm}(3.7183\times 10^{-15})$ \\
10 & $10^{-7}$ & 1 & -2.5964  & 2.6850  & -4.9210  & 8.5790  & $4.7985  \hspace{3mm}( 2.7747\times 10^{-5})$ & $1.2076\times 10^{-7} \hspace{3mm}(3.7138\times 10^{-15})$ \\
11 & $10^{-7}$ & 1 & -2.5943  & 2.6861  & -5.1229  & 8.9335  & $4.9986  \hspace{3mm}( 2.4236\times 10^{-5})$ & $1.2070\times 10^{-7} \hspace{3mm}(3.7094\times 10^{-15})$ \\
12 & $10^{-7}$ & 1 & -2.5922  & 2.6872  & -5.3246  & 9.2877  & $5.1986  \hspace{3mm}( 2.1286\times 10^{-5})$ & $1.2064\times 10^{-7} \hspace{3mm}(3.7047\times 10^{-15})$ \\
13 & $10^{-7}$ & 1 & -2.5902  & 2.6883  & -5.5261  & 9.6417 & $5.3987  \hspace{3mm}( 1.8791\times 10^{-5})$ & $1.2058\times 10^{-7} \hspace{3mm}(3.7013\times 10^{-15})$ \\
14 & $10^{-7}$ & 1 & -2.5881  & 2.6894  & -5.7273  & 9.9955 & $5.5988  \hspace{3mm}( 1.6665\times 10^{-5})$ & $1.2052\times 10^{-7} \hspace{3mm}(3.6969\times 10^{-15})$ \\
15 & $10^{-7}$ & 1 & -2.5860  & 2.6905  & -5.9282  & 10.349 & $5.7988  \hspace{3mm}( 1.4844\times 10^{-5})$ & $1.2047\times 10^{-7} \hspace{3mm}(3.6916\times 10^{-15})$ \\
16 & $10^{-7}$ & 1 & -2.5840  & 2.6916  & -6.1290  & 10.702 & $5.9989  \hspace{3mm}( 1.3274\times 10^{-5})$ & $1.2040\times 10^{-7} \hspace{3mm}(3.6874\times 10^{-15})$ \\
17 & $10^{-3}$ & 5 & $-4.0528\times 10^{3}$ & $6.9568\times 10^{4}$ & $-2.7455\times 10^5$ & $5.2670\times 10^5$ & $3.3968 \hspace{3mm} (3.3889\times 10^{-4}$) & $1.0252\times 10^{-3} \hspace{3mm}(2.4813\times 10^{-7})$ \\
% 18 & $10^{-3}$ & 5 & $-3.9198\times 10^{3}$ & $7.1002\times 10^{4}$ & $-2.4088\times 10^5$ & $5.8830\times 10^5$ & $<a_2>=7.2509 \hspace{3mm} (8.5876 \times 10^{-4}$) & $1.0138 \times 10^{-3} \hspace{3mm} 2.4088\times 10^{-7}$ \\
18 & $10^{-3}$ & 5 & $-3.9198\times 10^{3}$ & $7.1002\times 10^{4}$ & $-2.4088\times 10^5$ & $5.8830\times 10^5$ & $\tablenotemark{a} <a_2>=7.2509 \hspace{3mm} (8.5876 \times 10^{-4}$) & $1.0138 \times 10^{-3} \hspace{3mm} (2.4088\times 10^{-7})$ \\
 & & & & & & & \multicolumn{1}{r}{ ,$<a_3>=9.4989 \hspace{3mm} (1.1202 \times 10^{-3}$)} & \\
\enddata
\tablecomments{\scriptsize{
$a_1$ : the semi-major axis of the innermost planet. $E_{tot}$ : 
the total energy. $L_{ztot}$ : the z-component of the total angular
    momentum. $P_{xtot}, \, P_{ytot}$ : the x- and y-components of the
    total linear momentum.
$<\Delta>, \, \sigma^2_{\Delta}$ : the mean and variance of the
    planetary orbital separations over $10000$ realizations.
$<e>, \, \sigma^2_{e}$ : the mean and variance of the planetary
    eccentricity over $10000$ realizations. For each realization 
    the mean eccentricity over three planets, $e$, is obtained and 
    then $<e>$ and $\sigma^2_{e}$ are computed.}
}
\tablenotetext{a}{\scriptsize{
$<a_2>, \, \sigma^2 _{a_2}, \, <a_3>, \, \sigma^2_{a_3}$ : the mean
    and variance of the semi-major axis of the second and third planets. 
}}
\label{models}
\end{deluxetable} %
%%%%%%%%%%%%%%%%%%%%%%%%%%%%%%%%%%%%%%%%%%%%%%%%%%%%%%%%%%%%%%%%%%%%%%%
\begin{deluxetable}{cccccccc}
% \tabletypesize{\scriptsize}
\tabletypesize{\footnotesize}
\tablecaption{The Power Spectral Index for the Proto-Planet Systems}
\tablewidth{0pt}
\tablehead{
\colhead{Model} & \colhead{Orbital separation} & \multicolumn{6}{c}{power-law index $\nu$}\\ \hline
& \colhead{} & \multicolumn{2}{c}{sample 1}& \multicolumn{2}{c}{sample 2} & \multicolumn{2}{c}{sample 3}\\
 & \colhead{$\Delta$(Hill radius)} &  \colhead{1st phase} & \colhead{2nd phase} & \colhead{1st phase} & \colhead{2nd phase} & \colhead{1st phase} & \colhead{2nd phase}
}
\startdata
1 &2.9970 & - & 2.12 & - & 1.85 & - & 1.95 \\
2 &3.1973 & - & 1.59 & - & 2.06 & - & 1.36 \\
3 &3.3975 & - & 1.61 & - & 2.38 & - & 1.79 \\
4 &3.5977 & - & 1.83 & - & 1.59 & - & 2.03 \\
5 &3.7979 & - & 2.00 & - & 1.70 & - & 1.43 \\
6 &3.9980 & - & 2.03 & - & 1.52 & - & 1.90 \\
7 &4.1982 & - & 1.91 & - & 2.00 & - & 1.73 \\
8 &4.3983 & - & 1.66 & - & 2.01 & - & 1.79 \\
9 &4.5984 & - & 1.67 & - & 1.66 & - & 2.08 \\
10 &4.7985 & 0.10 & 1.89 & 0.65 & 2.01 & 1.27 & 1.39 \\
11 &4.9986 & - & 2.12 & 1.01 & 1.85 & - & 1.95 \\
12 &5.1986 & 0.75 & 1.62 & - & 1.92 & 0.91 & 2.31 \\
13 &5.3987 & 0.83 & 0.93 & 0.93 & 1.90 & 0.68 & 1.96 \\
14 &5.5988 & 0.84 & 1.85 & 1.25 & 1.97 & 0.85 & 1.96 \\
15 &5.7988 & 0.92 & 1.88 & 0.92 & 1.93 & 0.50 & 1.86 \\
16 &5.9989 & 1.09 & 1.80 & 0.65 & 1.84 & 0.90 & 1.95
\enddata
\tablecomments{
The power spectra of the orbital eccentricity for the innermost planet
are fit by the power law. Three realizations are arbitrarily chosen
among $10000$ realizations for each model and shown as samples 1, 2
and 3. The pre- and post-encounter phases are treated separately and
referred to as 1st and 2nd phases in the table, respectively.
For some of the models, the duration of the phase is too short to
obtain the power-law index and "-" is put instead of the index.
The averages of $\nu$ over all the models in this table are $0.84$ 
before the encounter and $1.84$ after the encounter.} 
\label{spsmall}
\end{deluxetable} 

%%%%%%%%%%%%%%%%%%%%%%%%%%%%%%%%%%%%%%%%%%%%%%%%%%%%%%%%%%%%%%%%%%%%%%%
\begin{deluxetable}{cccccccc}
% \tabletypesize{\scriptsize}
\tabletypesize{\scriptsize}
\tablecaption{The Power Spectral Index for the Jupiter Systems}
\tablewidth{0pt}
\tablehead{
\colhead{Model} &
\colhead{Orbital separation} & \multicolumn{6}{c}{power-law index $\nu$}\\ \hline
& \colhead{} & \multicolumn{2}{c}{sample 1}& \multicolumn{2}{c}{sample 2} & \multicolumn{2}{c}{sample 3}\\
& \colhead{} &  \colhead{1st phase} & \colhead{2nd phase} & \colhead{1st phase} & \colhead{2nd phase} & \colhead{1st phase} & \colhead{2nd phase}
}
\clearpage
\startdata
 1 & $<\Delta>=3.3968$(Hill radius) & 0.88 & 1.85 & 1.69 & 1.55 & 0.00 & 1.72 \\
2 & {\scriptsize $<a_2>=7.2509,<a_3>=9.4989$(AU)} & 0.99 & 1.87 & 1.03 & 1.78 & 0.99 & 1.87
\enddata
\tablecomments{The power spectra of the orbital eccentricity for the
  innermost planet are fit by the power law. The averages of $\nu$
  over the models in the table are $0.91$ before the encounter and
  $1.77$ after the encounter.}
\label{spbig}
\end{deluxetable} %
%%%%%%%%%%%%%%%%%%%%%%%%%%%%%%%%%%%%%%%%%%%%%%%%%%%%%%%%%%%%%%%%%%%%%%%

\end{document}